\documentclass[twocolumn,trackchanges]{aastex631}
\usepackage{todonotes}

\begin{document}

\title{Multiwavelength Evidence for Two New Candidate Transitional Millisecond Pulsars in the Sub-luminous Disk State: 4FGL J0639.1--8009 and 4FGL J1824.2+1231}

\author[0009-0006-2411-5162]{Rebecca Kyer}
\affiliation{Center for Data Intensive and Time Domain Astronomy, Department of Physics and Astronomy, Michigan State University, East Lansing, MI 48824, USA}

\author[0009-0005-5402-8900]{Subhroja Roy}
\affiliation{Center for Data Intensive and Time Domain Astronomy, Department of Physics and Astronomy, Michigan State University, East Lansing, MI 48824, USA}

\author[0000-0002-1468-9668]{Jay Strader}
\affiliation{Center for Data Intensive and Time Domain Astronomy, Department of Physics and Astronomy, Michigan State University, East Lansing, MI 48824, USA}

\author[0000-0003-1814-8620]{Ryan Urquhart}
\affiliation{Center for Data Intensive and Time Domain Astronomy, Department of Physics and Astronomy, Michigan State University, East Lansing, MI 48824, USA}

\author[0009-0004-4418-0645]{Isabella Molina}
\affiliation{Center for Data Intensive and Time Domain Astronomy, Department of Physics and Astronomy, Michigan State University, East Lansing, MI 48824, USA}

\author[0000-0002-3673-0668]{Peter Craig}
\affiliation{Center for Data Intensive and Time Domain Astronomy, Department of Physics and Astronomy, Michigan State University, East Lansing, MI 48824, USA}

\author[0000-0002-8400-3705]{Laura Chomiuk}
\affiliation{Center for Data Intensive and Time Domain Astronomy, Department of Physics and Astronomy, Michigan State University, East Lansing, MI 48824, USA}

\submitjournal{ApJ}
\accepted{2025 March 7}

\begin{abstract}


We report the discovery of two new Galactic accreting compact objects consistent with the respective positions of the unassociated Fermi-LAT $\gamma$-ray sources 4FGL J0639.1--8009 and 4FGL J1824.2+1231. A combination of new and archival X-ray data from Chandra, XMM-Newton, Swift/XRT, and eROSITA reveals a variable X-ray source in each $\gamma$-ray error ellipse. Both candidate counterparts show power-law spectra with photon indices $\Gamma \sim 1.7-1.9$. Optical follow-up photometry and spectroscopy show rapid high-amplitude variability unrelated to orbital motion and persistent accretion disk spectra for both objects. We demonstrate that the properties of these X-ray/optical sources are at odds with the known phenomenology of accreting white dwarfs, but are consistent with the observed properties of the sub-luminous disk state of transitional millisecond pulsars. This brings the census of confirmed or candidate transitional millisecond pulsars in the Galactic field to nine. We show this potentially represents $\lesssim 10\%$ of the total population of transitional millisecond pulsars within 8 kpc.

\end{abstract}

\section{Introduction} \label{sec:intro}

Transitional millisecond pulsars (tMSPs) are a rare class of binary neutron stars that transition between a rotation-powered radio pulsar state and an accretion-powered disk state \citep[see reviews by][]{Campana_diSalvo2018, Papitto_deMartino2022}). tMSPs give insight into the end stages of millisecond pulsar (MSP) recycling, whereby neutron stars are spun up to their rapid rotation speeds by accreting angular momentum from a companion over Gyr timescales \citep{Alpar1982, Bhattacharya1991}. 

Three tMSPs have been confirmed so far: PSR J1023+0038 \citep{Archibald2009, Wang2009} and XSS J12270-4859 \citep{Bassa2014,Roy2015} in the Galactic field, and PSR J1824–2452I in the globular cluster M28 \citep{M28I_Papitto2013}. While in the radio pulsar state, these systems appear as ordinary redback spider millisecond pulsars with variable X-ray emission attributed to an intrabinary shock \citep{vanderMerwe2020_spider_shocks} and modulated optical emission from the irradiated surface of the tidally-locked companion \citep[companion mass $M \gtrsim 0.1 M_{\odot}$,][]{Strader2019_redback_demographics, Stringer2021}. 

In their accretion-powered state, tMSPs display a unique combination of multiwavelength signatures linked to instabilities in the accretion disk and ejections of material \citep[e.g.,][]{Linares2022, J1023_Baglio2023}. Disk-state tMSPs are several times brighter in the $\gamma$-ray regime than when in the rotation-powered state \citep{Stappers2014, Torres2017} arising from an unknown mechanism in addition to the fraction of their spindown luminosity emitted as $\gamma$-rays that is seen in normal pulsars \citep{Abdo2013}. At X-ray energies, rapid switches between high (of order $L_X \sim 10^{33}$ erg s$^{-1}$) and low (of order $L_X \sim 10^{32}$ erg s$^{-1}$) emission modes are seen on timescales of seconds lasting for minutes, along with sporadic flares an order of magnitude brighter \citep[e.g.,][]{J1023_Baglio2023}. The typical X-ray luminosity is lower than observed for other neutron star low-mass X-ray binaries (NS-LMXBs) which have X-ray luminosities during accretion episodes in the range $L_X \sim 10^{35}-10^{38}$ erg s$^{-1}$ \citep[see compact binary radio/X-ray luminosity correlation database compiled by][]{Bahramian_Rushton}. This ``sub-luminous" disk state suggests that mass transfer in tMSPs may occur at the lowest rates observed in NS-LMXBs \citep{Linares2022}, or at lower radiative efficiency \citep{rad_efficiency:qiao_liu:2021}. The hot accretion disk dominates the optical spectra of tMSPs, as seen for other LMXBs, and strong optical variability occurs on timescales of minutes, possibly arising from the same accretion flow instabilities in the inner disk that cause the X-ray moding. At the longest wavelengths, pulsed radio emission is not detected and disk-state tMSPs instead show radio continuum emission. In PSR J1023+0038 the radio emission has been found to be anti-correlated to the X-ray modes, with radio flares during low X-ray modes that suggest a disruption of the accretion flow due to mass ejections \citep{J1023_Baglio2023}, in addition to persistent continuum emission that may be from a self-absorbed synchrotron jet \citep{J1023_Deller2015}. Simultaneous X-ray and radio observations of the candidate 3FGL J0427.9–6704 also show evidence for a radio jet or continuous outflow, without the presence of X-ray moding \citep{J0427_Li2020}.

The physical processes that produce the complex phenomenology of tMSPs have yet to be understood. Few studies have attempted to simulate the interactions between the pulsar wind, companion, and accretion disk to reproduce the observed variability and transition behavior \citep{Parfrey2017, Murguia-Berthier2024, Guerra2024}. Multi-dimensional magnetohydrodynamic simulations are likely necessary to fully study these systems, therefor observationally constraining the range of physical parameters to consider is needed to make these simulations feasible.

The prototypical tMSP PSR J1023+0038 has been the best-observed of the three confirmed transitional sources owing to its close distance of 1.4 kpc. Whether the observed signatures of the disk state described above occur in all tMSPs remains to be confirmed by discovery of a larger tMSP sample. A handful of candidate tMSPs in the Galactic field that have properties consistent with the phenomenology of the accretion-powered state have been put forth in the literature (e.g., 3FGL J1544.6–1125, \cite{J1544_discovery}; 3FGL J0427.9–6704, \cite{J0427_Strader2016, J0427_Li2020}; CXOU J110926.4–650224, \cite{J1109_discovery}; 4FGL J0407.7-5702, \cite{J0407_discovery}; 4FGL J0540.0-7552, \cite{J0540_discovery}). Like the confirmed tMSPs, these candidates display optical spectra dominated by an accretion disk, strong X-ray variability, and are associated with $\gamma$-ray sources. 

Of the three confirmed tMSPs, only PSR J1023+0038 is currently in the accretion-powered state, so the candidate systems offer a necessary avenue to study the range of properties disk-state tMSPs might display. Identifying candidate tMSP systems may also reveal eclipsing systems, such as 3FGL J0427.9-6704, which can offer strong constraints on the masses, radio outflow geometry, and other properties while in the disk state. Discovering more candidate tMSP systems is therefore a high priority in furthering our understanding of these unusual systems.

As part of an ongoing program to obtain optical spectroscopy of the counterparts of X-ray sources located within unassociated Fermi Large Area Telescope (LAT) sources, we created a spatial cross match between the 4FGL-DR4 catalog and several X-ray catalogs using the Bayesian cross matching tool NWAY \citep{NWAY}. We present the discovery of new accreting compact binary systems within the 68\% error ellipse of two unassociated Fermi $\gamma$-ray sources, 4FGL J0639.1-8009 and 4FGL J1824.2+1231, and show that the properties of each source's proposed multiwavelength counterparts are consistent with those of the tMSP sub-luminous disk state.

\section{Data} \label{sec:data}
\subsection{$\gamma$-Ray}
4FGL J0639.1-8009 and 4FGL J1824.2+1231 are unassociated sources that were first included in the incremental 12 yr 4FGL DR3 and 8 yr 4FGL DR1 catalogs, respectively \citep{4FGL-DR3}. In the most recent 14 yr 4FGL DR4 catalog \citep{4FGL-DR4}, the sources have positional error ellipses with semi-major axes of $7.6\arcmin$ and $5.6\arcmin$ at the 68\% confidence level, and $12.3\arcmin$ and $9.0\arcmin$ at the 95\% confidence level. Both sources show evidence for a curved spectral shape, with the LogParabola model being a $\sim 2\sigma$ improvement over the PowerLaw fit to each observed spectrum. The observed fluxes in the $0.1-100.0$ GeV energy band are $(2.0 \pm 0.5) \times 10^{-12}$ erg s$^{-1}$ cm$^{-2}$ and $F_{\gamma} = (2.7\pm 0.9)\times 10^{-12}$ erg s$^{-1}$ cm$^{-2}$. The variability index listed in the 4FGL DR4 catalog for each source is 22.1 and 14.8, which is lower than the value (27.7) at which sources are formally classified as being variable in 4FGL DR4. Future data releases will allow an improved assessment of variability.

\subsection{X-Ray \label{sec:data:xray}} 

The brightest X-ray source in the 68\% positional uncertainty of 4FGL J0639.1-8009, which we will argue is the most likely counterpart to the $\gamma$-ray source, has been previously detected in observations from five different X-ray observatories between 2013 and 2017 during targeted observations of nearby sources. In Table \ref{tab:xray_counterparts} we give a summary of the catalog detections, including shorter exposure detections by the ROSAT and eROSITA all-sky surveys. We also consider the second-brightest X-ray source within the 68\% error ellipse of 4FGL J0639.1-8009 as one that could alternatively be the source of the $\gamma$-ray emission. Only one X-ray source has been detected in the 68\% positional uncertainty of 4FGL J1824.2+1231, by Swift/XRT. Except for the XMM-Newton observation, all X-ray spectral fits were performed with XSPEC version 12.14.0. The results of all new spectral fits are detailed in Section \ref{sec:results:xray:spectra}.

\subsubsection{XMM-Newton Targeted Observation}

The source within 4FGL J0639.1-8009 detected by XMM lies $5.2\arcmin$ from the center of the Fermi error ellipse and was observed on 2016 May 3 in a 47 ksec observation (0781710101, PI: G. Israel). The source was given the designation 4XMM J064059.5-801125 in the 4XMM-DR9 Serendipitous Source Catalog \citep{XMM}, which re-analyzed all public XMM observations obtained through 2019. The source effective exposure time was $\sim23$ ksec after filtering periods of high background flaring. The mean net EPIC count rate was $(1.9 \pm 0.1) \times 10^{-1}$ ct s$^{-1}$ from 966 net source counts. Due to the source's position on a chip gap, the expected PSF-weighted detector coverage on the pn chip (PN\_MASKFRAC) is only 39\%.

We reprocessed the XMM data using standard tasks within the Science Analysis System \citep[SAS;][]{SAS} version 18.0.0 software package. As these data were not designed to target 4FGL J0639.1-8009, the X-ray source of interest is significantly off-axis ($\sim11\arcmin$ from the pointing center) and lies outside of the chip for the MOS1 detector. For MOS2, we used a circular source extraction region with radius of $30\arcsec$, and local background region three times larger. As the target is partially obstructed by a chip gap in the pn image, we use a rectangular source extraction region, with a similarly large local background, taking care to avoid the chip gap. Intervals of high particle background flaring were filtered out. Standard flagging tools were applied to both the MOS2 (\verb|#XMMEA_EM|) and pn (\verb|#XMM_EP && FLAG==0|) data, and we use the patterns 0-12 and 0-4, respectively.

Background-subtracted light curves for each camera were created using the SAS tasks \textit{evselect} and \textit{epiclccorr}. Barycentric corrections were applied using the \textit{barycen} task. Finally, the MOS2 and pn light curves were combined using the FTOOLS \citep{FTOOLS} command \textit{lcmath}. Individual background-subtracted spectra were created using \textit{xmmselect} and then combined with \textit{epicspeccombine}. The final combined spectrum was binned to at least 20 counts per bin so that Gaussian statistics could be used for spectral fitting, performed using XSPEC \citep{xspec} version 12.10.1.

\subsubsection{Swift Observations}

The position of the XMM detection was also observed by Swift for 47 ksec total exposure time across 16 observations between Dec 2016 --- Aug 2017, during monitoring of the same unrelated source that the XMM observation was targeting. In the Living Swift X-ray Point Source catalog \citep{LSXPS}, the observations have a combined signal to noise ratio ($S/N$) $\sim 23$ for 374 net source counts in the 0.3---10.0 keV band, with a mean count rate of $(9.7 \pm 0.5) \times 10^{-3}$ ct s$^{-1}$. 

The only X-ray source in the positional uncertainty of 4FGL J1824.2+1231 was LSXPS J182408.7+123231 with 29 net source counts in the 0.3---10.0 keV band, observed for 3.5 ksec across 27 snapshots between Aug 2016---Nov 2019. In addition to this archival detection, we obtained an 8.5 ksec unanticipated Target of Opportunity observation with Swift/XRT in photon counting mode (ToO ID 21032) on 2024 August 31 to better constrain the model fit to the X-ray spectrum. A total of 136 net source counts were observed in the combined 11 ksec exposure time with mean $S/N = 9.5$ and the $90\%$ positional uncertainty was reduced to $3.8\arcsec$. We reduced the observations of both sources using the online XRT product tools \citep{Evans2020}.

\subsubsection{Chandra Targeted Observation}

The X-ray source we propose as the counterpart to 4FGL J0639.1-8009 was detected by the Chandra X-ray Observatory on 2013 August 27 in a 38.5 ksec observation targeting another source (ObsID 14925; PI: A. Wolter). The source lies about $7\arcmin$ off-axis from the pointing center on the edge of an ACIS-I detector used in the VFAINT mode, close enough to the edge of the CCD that the source is not included in the Chandra Source Catalog \citep[CSC v2.1;][]{CSC2.1}, and therefor is not included in Table \ref{tab:xray_counterparts}. We reprocessed this observation with the \textit{chandra\_repro} script in CIAO version 4.16 \citep{ciao} and CALDB version 4.11.3. We extracted 449.1 net source counts in this observation for a mean count rate of 0.0117 ct s$^{-1}$. 

\subsubsection{eROSITA and ROSAT Survey Observations}
The first X-ray detection at the position of the 4FGL J0639.1-8009 counterpart was observed by the Position Sensitive Proportional Counter on board the ROSAT observatory in 1990 with an exposure time of 0.5 ksec. Its catalog entry in the Second ROSAT All-Sky Survey Point Source Catalog \citep[2RXS;][]{ROSAT} has 15.8 net source counts in the 0.1---2.4 keV energy band, and mean background-corrected count rate $(3.2 \pm 1.0) \times 10^{-2}$ ct s$^{-1}$. The detection likelihood of 16.6 indicates a spurious detection probability of $< 1\%$. We converted the 2RXS count rate to flux with WebPIMMS assuming an absorbed power law spectrum model with $\Gamma = 1.7$ for the entry in Table \ref{tab:xray_counterparts}. 

Most recently, an X-ray detection consistent with the position of the counterparts to 4FGL J0639.1-8009 proposed here is included in the first public eROSITA All-Sky Survey data release \citep[1eRASS;][]{1eRASS} with $\sim9\sigma$ significance in both the main 0.2---2.3 keV catalog and the hard 2.3---5 keV catalog. The source was observed between 2020 March 22 and 2020 March 26 for 0.44 ksec total exposure time. About 102 net source counts were detected in the 0.2---5.0 keV band, for a mean count rate of $(2.4 \pm 0.3) \times 10^{-2}$ ct s$^{-1}$.

\subsubsection{Other X-ray Sources Within the $ 68\%$ $\gamma$-Ray Error Ellipses}

We checked for additional bright X-ray sources within the several-arcminute Fermi positional uncertainties that could belong to a source class known to produce $\gamma$-rays. There are archival observations at various depths by XMM, Swift, Chandra and eROSITA within the field of 4FGL J0639.1-8009. In all catalogs, the source that we propose as the true counterpart to the $\gamma$-ray source is the brightest X-ray source in the 68\% uncertainty ellipse by a factor of at least four. The next brightest source is a likely active galactic nucleus \citep[AGN;][]{WISEAGN, ML_4XMMDR9, ML_Chandra} located $5.7\arcmin$ from the center of the Fermi detection, slightly further from the center than our source, and included in catalogs by XMM (4XMM J063720.2-801230), Swift (LSXPS J063719.9-801229) and Chandra (2CXO J063719.7-801230). We will argue that this AGN is not associated with the $\gamma$-ray source in Section \ref{sec:discussion:agn}. All other archival X-ray sources in the Fermi ellipse of 4FGL J0639.1-8009 are even fainter, and we did not find any other classified sources that are likely $\gamma$-ray emitters. In the field of 4FGL J1824.2+1231, the source detected by Swift is currently the only known X-ray source in the 68\% error ellipse.

\begin{deluxetable*}{lllccc}
\tablewidth{1.0\textwidth}
\tablehead{
    \colhead{Catalog Name} & \colhead{R.A.} & \colhead{Decl.} & \colhead{Pos. uncertainty} & \colhead{Energy band} & \colhead{Mean flux} \\
    \colhead{} &\colhead{(J2000)} & \colhead{(J2000)} & \colhead{} & \colhead{} & \colhead{$10^{-13}$ erg s$^{-1}$ cm$^{-2}$}}

\tablecaption{Summary of catalog detections of high-energy counterparts to 4FGL J0639.1-8009 and 4FGL J1824.2+1231 proposed here. Positional uncertainty is quoted at different confidence levels by each catalog, indicated in parentheses next to each entry, and the semi-major axis is reported for the Fermi sources. Absorbed fluxes are given with 1-sigma uncertainties as reported in each catalog, except the 2RXS flux which comes from converting the mean count rate assuming a power law flux model with $\Gamma=1.7$. An additional detection consistent with the positions corresponding to 4FGL J0639.1-8009 is included in a Chandra observation of the field, but it lies close to the edge of the CCD it was detected on so is not present in the Chandra Source Catalog. \label{tab:xray_counterparts}}
\startdata
4FGL J0639.1-8009 & 06:39:09 & --80:09:10 & $7.6\arcmin$ (68\%) & 0.1---100 GeV &  $21 \pm 5$ \\ 
4XMM J064059.5-801125 & 06:40:59.5 & --80:11:25.7 & $1.0\arcsec$ (68\%) & 0.2---12.0 keV  & $6.1 \pm 0.6$ \\ 
LSXPS J064059.4-801127 & 06:40:59.4 & --80:11:27.2 & $3.6\arcsec$ (90\%) & 0.3---10.0 keV & $4.0 \pm 0.2$ \\
1eRASS J064100.6-801127 & 06:41:00.6 & --80:11:27.8 & $1.6\arcsec$ (68\%) & 0.2---5.0 keV & $4.0 \pm 0.9$\\
2RXS J064103.6-801134 & 06:41:03.6 & --80:11:34.3 & $31.7\arcsec$ (90\%) & 0.1---2.4 keV & $4.6 \pm 1.4$ \\ 
\hline
4FGL J1824.2+1231 & 18:24:16 & +12:31:49 & $5.6\arcmin$ (68\%) & 0.1---100 GeV & $27 \pm 9$ \\
LSXPS J182408.7+123231 & 18:24:08.79 & +12:32:31.4 & $3.5\arcsec$ (90\%) & 0.3---10.0 keV & $4.3 \pm 0.4$
\enddata
\end{deluxetable*}

\subsection{Optical}
\subsubsection{Gaia}

We identified unique Gaia DR3 optical counterparts to the X-ray sources proposed as the counterparts to both $\gamma$-ray sources. For 4FGL J0639.1-8009, Gaia DR3 5207836863615934080 lies within the positional uncertainty of the Swift and XMM X-ray sources located at the ICRS epoch 2016 position (R.A., Decl.) = (06:40:59.501, --80:11:26.189). We will refer to this source as J0640A. J0640A is faint ($G \sim 20.6$), quite blue ($Bp-Rp \sim 0.9$) and has high proper motion $\mu = 15.5 \pm 1.9$ mas yr$^{-1}$ but a weak bias-corrected parallax measurement of $\varpi = 0.72 \pm 1.04$ \citep{Lindgren2021}, consistent with either a distant Galactic object or potentially much closer. 

For 4FGL J1824.2+123, Gaia DR3 4485064648767923328 is the only optical source within the positional uncertainty of LSXPS J182408.7+123231, located at the ICRS epoch 2016 position (R.A., Decl.) = (18:24:8.884, +12:32:33.314). We will refer to this source as J1824A. The $G\sim 19.3$ source is somewhat blue ($Bp-Rp \sim 1.4$) and has high proper motion $\mu = 18.5 \pm 0.4$ mas yr$^{-1}$. With a bias-corrected parallax of $\varpi = 0.66 \pm 0.27$ mas \citep{Lindgren2021}, the median distance is 1.5 kpc with a $2\sigma$ range of 0.8--6.4 kpc. This corresponds to a transverse velocity of $\sim 130$ km s$^{-1}$ at 1.5 kpc, but even higher at the larger distances. We discuss other distance estimates and the likely distance range to both sources in Section \ref{subsec:distances}.

\subsubsection{SOAR/Goodman Spectroscopy}
We obtained 15 optical spectra of J0640A with the Goodman Spectrograph \citep{goodman} on the SOAR telescope located on Cerro Pach\'{o}n in Chile on four nights between 2024 February 1 and 2024 March 11, and 17 optical spectra of J1824A with SOAR/Goodman on three nights between 2024 July 2 and 2024 August 10. These spectra were taken with 25 minute integration times using a 400 line mm$^{-1}$ grating and a 1.2\arcsec\ slit, resulting in a mean resolution of 7.2 \AA\ over a wavelength range 4000--7800 \AA. When J0640A was again visible at SOAR, we obtained four additional medium-resolution spectra of J0640A on 2024 September 10 with 30 minute integration times and the 930 line mm$^{-1}$ grating with the 1.2\arcsec\ slit, resulting in mean resolution 3.2 \AA\ over the wavelength range 5500--7200 \AA. Two more spectra of J0640A were obtained on 2024 Nov 3 and four more on 2024 Nov 25 with 30 minute integration times using the 400 line mm$^{-1}$ grating and a 1.2\arcsec\ slit.

All spectra were reduced and optimally extracted using standard procedures in IRAF \citep{IRAF,Tody93,Fitzpatrick24}, then wavelength-calibrated using FeAr arcs taken after each set of science exposures. As described in detail in \citet{Dodge2024}, for the 400 line mm$^{-1}$ data, we corrected the zeropoint of the wavelength solution by fitting the telluric A band with a model created in {\tt TelFit} \citep{Gullikson2014}.

\subsubsection{SOAR/Goodman Imaging}
We obtained $i^{\prime}$-band imaging of J0640A on 2024 March 21 using 180 s exposures in the imaging mode of the Goodman Spectrograph over 2.2 hr. We obtained wide-band imaging of J1824A with 30 s exposures in the GG395 filter on 2024 July 7 with SOAR/Goodman over a baseline of almost 3 hr. We performed differential aperture photometry with respect to 19 and 31 nonvariable comparison stars within $2.5\arcmin$ of each target using standard tools in IRAF \citep{IRAF,Tody93,Fitzpatrick24}. Due to the similarity between the GG395 filter and the Gaia $G$ band, we calibrated our photometry of J1824A to the $G$ band. The photometric measurements are given in Table \ref{tab:soar_photometry} and Table \ref{tab:j1824_soar_photometry}. For the photometry of J0640A, we find a systematic uncertainty of 0.03 mag from the RMS of the absolute offsets between our measured magnitude and Skymapper Southern Sky Survey photometry of four test stars excluded from the comparison star sample. Given the known color-dependent offsets between the Skymapper and SDSS filters\footnote{\href{https://skymapper.anu.edu.au/filter-transformations/}{https://skymapper.anu.edu.au/filter-transformations/}}, we estimate a more conservative calibration uncertainty of 0.06 mag. For J1824A we find a calibration uncertainty of 0.1 mag between our measured magnitudes and the Gaia filter system based on five test stars. These calibration uncertainties are not included in the light curve figures or tables below.

\begin{deluxetable}{lcc}
\tablecaption{SOAR/Goodman photometry of J0640A. All dates are Barycentric Modified Julian Dates. \label{tab:soar_photometry}}
\tablewidth{1.0\columnwidth}
\tablehead{\colhead{BMJD} & \colhead{$i'$} & \colhead{Unc.} \\
            \colhead{} & \colhead{(mag)} & \colhead{(mag)}}
\startdata
60390.1464241 & 20.31 & 0.03 \\
60390.1486573 & 20.65 & 0.04 \\
60390.1508674 & 20.59 & 0.04 \\
60390.1534009 & 20.46 & 0.03 \\
60390.1555693 & 20.37 & 0.03 \\
60390.1577381 & 20.28 & 0.03 \\
60390.1599069 & 20.19 & 0.03 \\
60390.1620756 & 19.99 & 0.02 \\
60390.1642444 & 19.92 & 0.02 \\
60390.1664134 & 19.98 & 0.02
\enddata
\tablecomments{A portion of Table \ref{tab:soar_photometry} is shown here for guidance regarding its form and content. The full table is available online.}
\end{deluxetable}

\begin{deluxetable}{lcc}
\tablecaption{SOAR/Goodman photometry of J1824A. All dates are Barycentric Modified Julian Dates. \label{tab:j1824_soar_photometry}}
\tablewidth{1.0\columnwidth}
\tablehead{\colhead{BMJD} & \colhead{$G$} & \colhead{Unc.} \\
            \colhead{} & \colhead{(mag)} & \colhead{(mag)}}
\startdata
60500.1809295 & 18.91 & 0.02 \\
60500.1813616 & 18.91 & 0.02 \\
60500.1818941 & 19.01 & 0.01 \\
60500.1823268 & 18.86 & 0.01 \\
60500.1827593 & 18.60 & 0.01 \\
60500.1831920 & 18.66 & 0.02 \\
60500.1836247 & 18.55 & 0.01 \\
60500.1840574 & 18.74 & 0.01 \\
60500.1844902 & 18.94 & 0.01 \\
60500.1849227 & 19.14 & 0.02 
\enddata
\tablecomments{A portion of Table \ref{tab:j1824_soar_photometry} is shown here for guidance regarding its form and content. The full table is available online.}
\end{deluxetable}

\subsubsection{Pan-STARRS and ZTF Light Curves}

\begin{figure}[t!]
\includegraphics[width=0.49\textwidth]{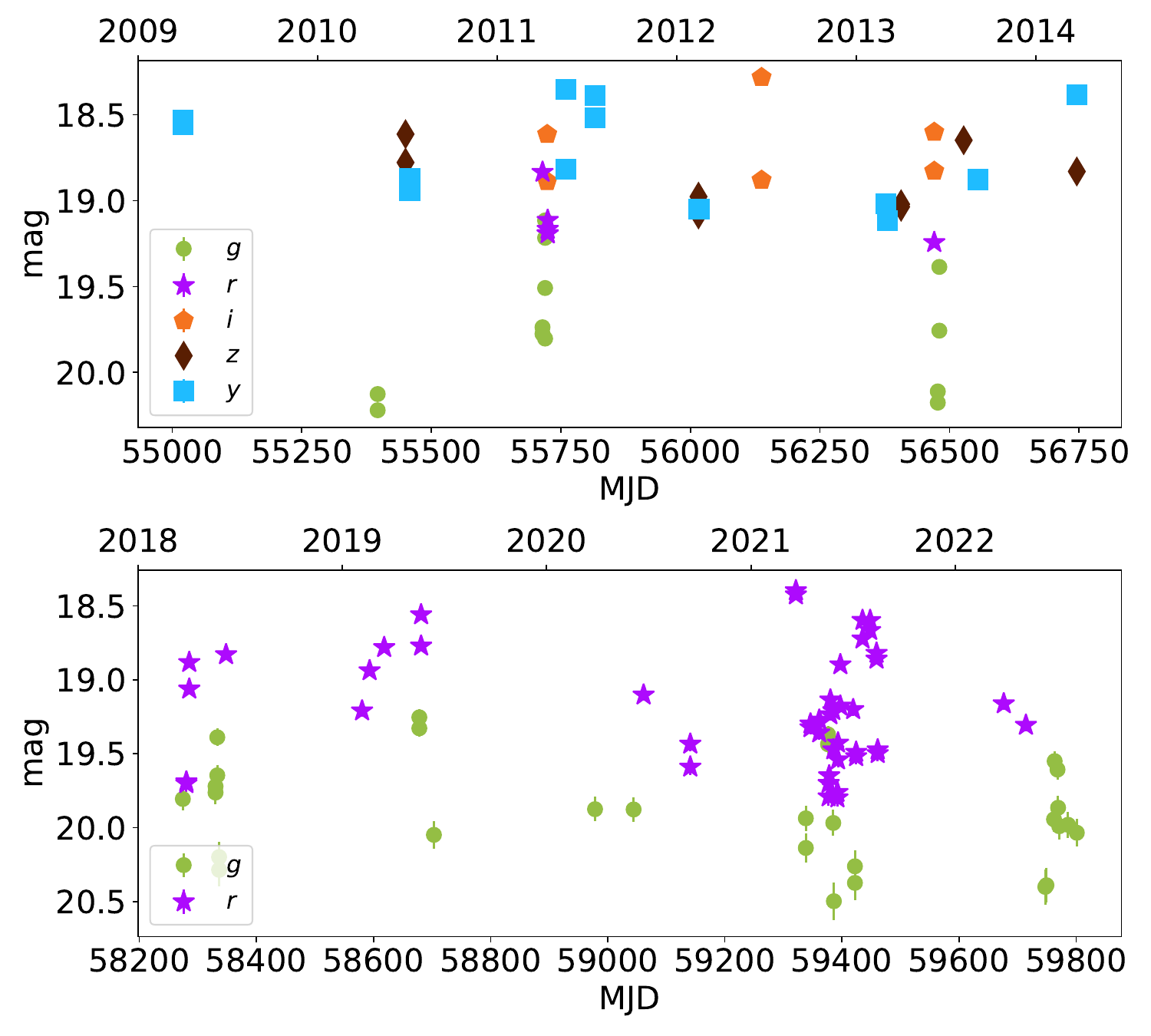}
\caption{Top panel: Pan-STARRS light curve of J1824A in the $grizy$ filters spanning 2009 to mid-2014; bottom panel: ZTF light curve of J1824A in the ZTF $g$ and $r$ filters spanning 2018 to late 2022. The error bars are smaller than the markers for most points. Both light curves show optical variability of order 0.5--1.0 mag in all available bands sustained over multiple years. \label{fig:ps_ztf_lc}}
\end{figure}

The field surrounding LSXPS J182408.7+123231 has been well observed by Northern Hemisphere optical surveys. J1824A was observed by the Panoramic Survey Telescope and Rapid Response System \citep[Pan-STARRS;][]{Chambers2016_panstarrs, Flewelling2020_panstarrs} between 2009 July 8 to 2014 March 29 and by the Zwicky Transient Facility \citep[ZTF;][]{Masci2019_ZTF} between 2018 June 5 to 2022 August 10. For the Pan-STARRS source (objID 123052760370011669) we only consider detections with the quality flag psfQfPerfect $> 0.95$. For the ZTF DR22 light curve (oid 1584103100034091 and 1584203100042589, corresponding to the $g$ and $r$ filters), we consider only epochs with catflags $< 32768$ to exclude poor-quality detections. We do not apply any color correction to the ZTF photometry to convert to the Pan-STARRS filter system, as the colors of these systems are highly variable. Both light curves, shown in Figure \ref{fig:ps_ztf_lc}, display significant amplitudes of variability in each band. We discuss the long-term variability details further in Section \ref{sec:results:opt_var}. 

\subsection{Radio}
The field of 4FGL J0639.1-8009 was imaged as part of the Rapid Australian Square Kilometer Array Pathfinder Continuum Survey (RACS) in 2021, 2022 and 2024 \citep{RACS1, RACS2}. The field around 4FGL J1824.2+1231 was imaged by RACS in 2021, 2022 and 2024. We analyzed these images with the beta web server version of the Cube Analysis and Rendering Tool for Astronomy \citep[CARTA;][]{CARTA}. No radio source is detected within the positional errors of the proposed X-ray counterparts to 4FGL J0639.1-8009 in Table \ref{tab:xray_counterparts}, down to a $3\sigma$ upper limit of 0.5 mJy in each RACS band (RACS-low at central frequency 944 MHz, RACS-mid at 1368 MHz and RACS-high at 1656 MHz). No radio source is detected at the position of the proposed X-ray counterpart to 4FGL J1824.2+1231 in any epoch, down to a $3\sigma$ upper limit of 0.2 mJy in the RACS-low band. For J1824A we also checked for radio coverage by Northern Hemisphere observatories, and obtained a $3\sigma$ upper limit of 0.4 mJy at the position in two Very Large Array Sky Survey images in 2019 and 2021 \citep{VLASS}.

\section{X-ray Results} \label{sec:results:xray}
\subsection{Mean X-ray Spectra and Fluxes \label{sec:results:xray:spectra}}

We fit the spectra of the XMM, Swift, Chandra and eROSITA observations of the counterparts that we proposed in Section \ref{sec:data:xray}, and searched for evidence of variability in J1824A. Due to the low counts, Cash statistics \citep{Cash} were used to determine goodness of fit for spectral models applied to all data sets except the XMM observation, which had sufficient counts in the combined pn+MOS2 spectrum to use Gaussian statistics. All new X-ray measurements are quoted with uncertainties at the 90\% confidence level. We find that the mean X-ray spectrum of J0640A is consistent within the errors of the fits across all observed epochs, from 2013--2020. The best fit model for both sources is a simple absorbed power law, with photon index $\Gamma = 1.7 \pm 0.2$ for J0640A and $\Gamma = 1.9 \pm 0.3$ for J1824A. We find that J1824A is likely variable, although more X-ray data is needed to rule out other spectral models and characterize said variability. 

\subsubsection{XMM Spectrum} 
For the X-ray source associated with J0640A, we fit the XMM/pn+MOS2 spectrum with multiple possible spectral models: an absorbed power law (\texttt{tbabs*powerlaw}), an absorbed thermal accretion disk (\texttt{tbabs*diskbb}), a composite model with both an absorbed power law and thermal component (\texttt{tbabs*[bbody+powerlaw]}), and an absorbed collisionally-ionized isothermal plasma model (\texttt{tbabs*apec}). The distance to J0640A is unknown but it lies at a high Galactic latitude of $b=-27.2^{\circ}$ so we attempted separate fits with the column density $N_H$ as a free parameter, and with a fixed $N_H = 1.12 \times 10^{21}$ cm$^{-2}$ corresponding to the full column density in the direction of the source given by the HI4PI Map \citep{HI4PI}. 

\begin{figure}[t!]
\includegraphics[width=1.0\columnwidth]{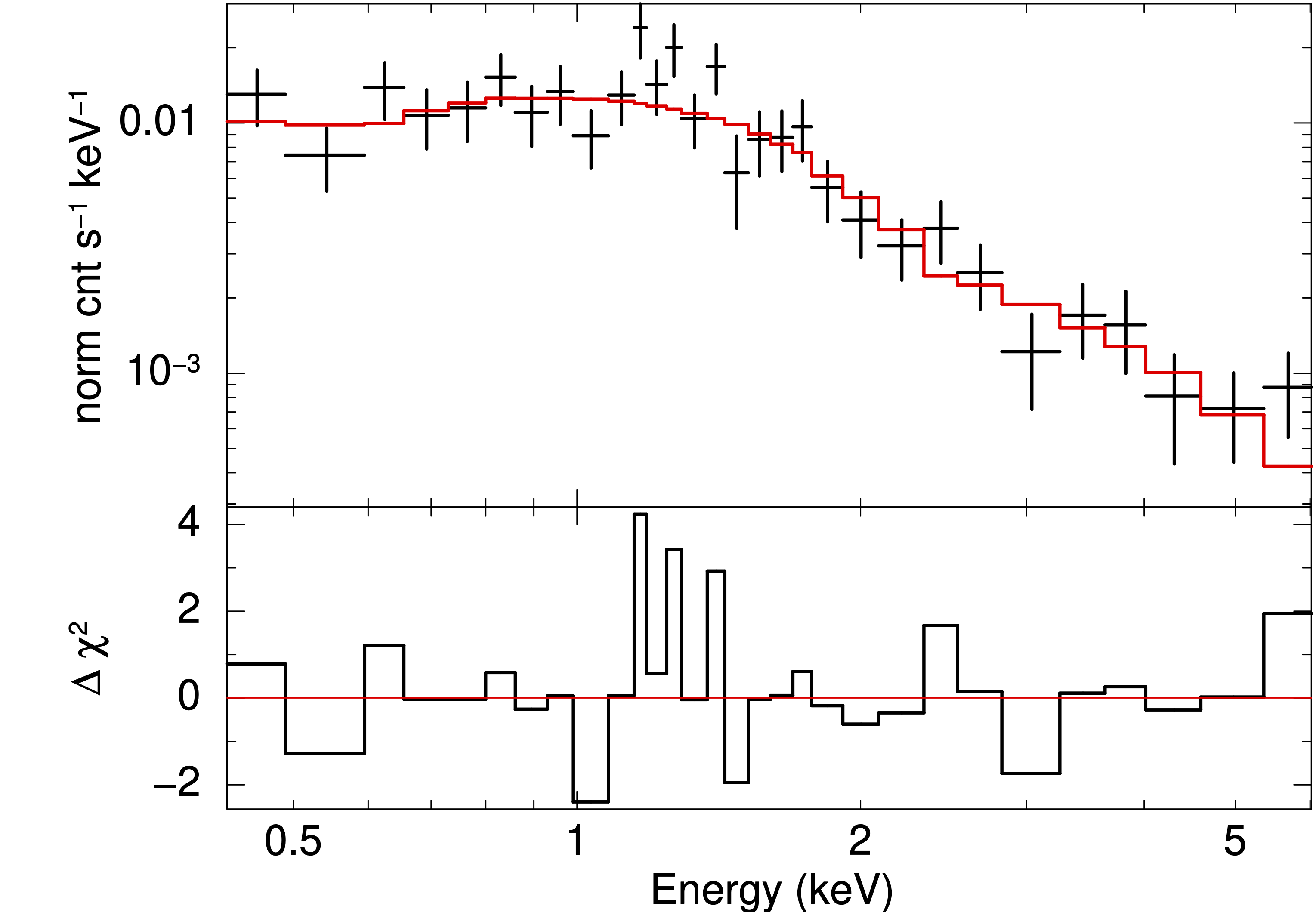}
\caption{Top panel: binned XMM spectrum of J0640A (black) and absorbed power law fit with $\Gamma = 1.8$ and $N_H = 1.2\times10^{21}$ cm$^{-2}$ (red); bottom panel: residuals. The absorbed power law model is a good fit to the data ($\chi^2$/d.o.f. = 27.8/27). \label{fig:xmm_spec}}
\end{figure}

With $N_H$ free, we found that a simple absorbed power law with photon index $\Gamma = 1.8^{+0.4}_{-0.3}$ and $N_H = (1.2^{+0.9}_{-0.8}) \times 10^{21}$ cm$^{-2}$ (consistent with the expected foreground $N_H$) is a good fit to the data ($\chi^2$/d.o.f. $=27.8/27$), shown in Figure \ref{fig:xmm_spec}. Fits of the two models with a blackbody component did not result in statistically significant improvements over the absorbed power law. The absorbed \texttt{apec} model with fixed metal abundances relative to solar was an unacceptable fit whether $N_H$ was free or fixed ($\chi^2$ / d.o.f. = 133.9/27 and 234.2/28, respectively). The only statistically good fit of an \texttt{apec} model ($\chi^2$ / d.o.f. = 28.5 / 26) we could obtain was when we allowed the metal abundances to vary, in which the abundance was not able to converge on a meaningful value and the plasma temperature was highly uncertain ($kT = 6.6^{+7.8}_{-3.3}$ keV). We can conclude that an absorbed \texttt{apec} model is ruled out by this data set. We use an absorbed power law for all fits to the other X-ray observations of J0640A.

As the source lies on a chip gap of the pn camera in this observation, we checked for energy dependence of the PSF that might affect the results of fitting the combined spectrum by fitting the spectra detected by each instrument independently. With $N_H$ free, there was minor disagreement in the spectrum's slope at soft energies $<1$ keV. With $N_H$ fixed to the expected foreground absorption, the differences in the slope were not statistically significant ($\Gamma = 2.1\pm0.3$ for the pn spectrum vs $\Gamma = 1.7\pm0.2$ for the MOS2 spectrum). The results of the combined spectral fit are consistent with the spectra of the individual detectors.

To measure the flux in this observation, we considered only the MOS2 spectrum. With $N_H$ fixed to the expected foreground absorption, we obtained a best fit (c-stat/d.o.f. = 371.8/384) absorbed power law to the spectrum with $\Gamma = 1.7 \pm 0.2$, resulting in the 1.0---10.0 keV unabsorbed flux $F_X = (3.8^{+0.8}_{-0.7}) \times 10^{-13}$ erg s$^{-1}$ cm$^{-2}$. The 0.2-12 keV unabsorbed flux of the source with this same model agrees what that published in 4XMM-DR14 (see Table \ref{tab:xray_counterparts}). 

\subsubsection{Swift/XRT Spectra}
For J0640A, we fit the Swift/XRT spectrum with an absorbed power law model, which was the best-fitting model to the higher-signal XMM spectrum. The best-fit model (c-stat/d.o.f. = 187.9/216) gave a photon index $\Gamma = 1.5^{+0.3}_{-0.2}$ and unconstrained $N_H = (5^{+9}_{-5}) \times 10^{20}$ cm$^{-2}$. With $N_H$ fixed to the expected foreground value, we obtained an improved fit of $\Gamma = 1.6 \pm 0.2$ which resulted in an unabsorbed 1.0---10.0 keV flux of $F_X = (4.2^{+0.7}_{-0.6})\times 10^{-13}$ erg s$^{-1}$ cm$^{-2}$. Both the flux and spectral shape are consistent with the XMM observation.

\begin{figure}[ht!]
\includegraphics[width=1.1\columnwidth]{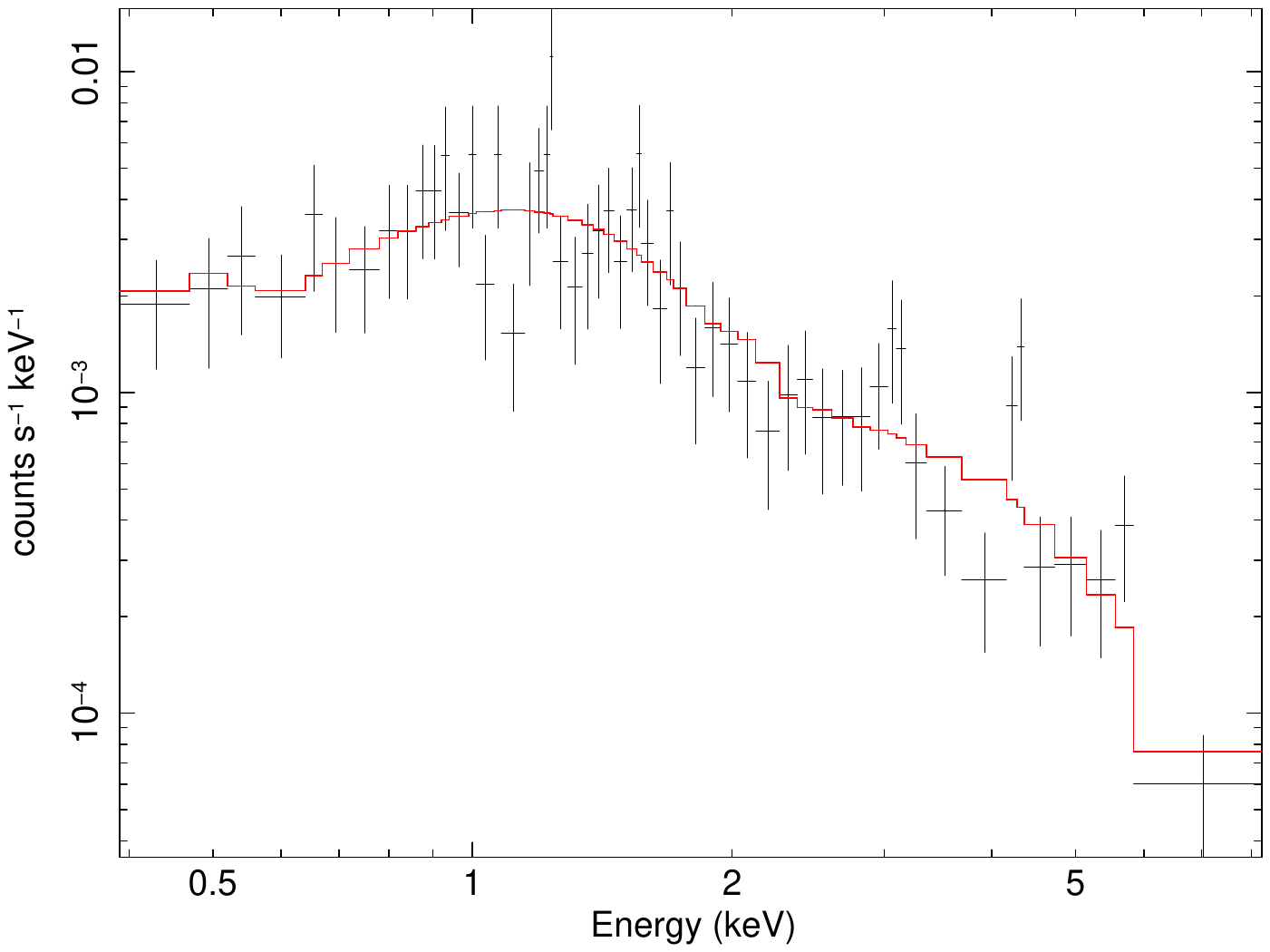}
\caption{Binned Swift spectrum of J0640A (black) and absorbed power law fit (red). The absorbed power law model is a good fit to the data (c-stat/d.o.f. = 187.9/216). \label{fig:swift_spec}}
\end{figure}

For J1824A, we found that a simple absorbed power law yielded a good fit to the observed Swift/XRT spectrum (c-stat/d.o.f. = 110.43/107). We fixed the Hydrogen column density to the expected foreground value of $N_H =1.43 \times 10^{21}$ cm$^{-2}$ and obtained a photon index $\Gamma = 1.86 \pm 0.25$, resulting in an unabsorbed 1.0---10.0 keV flux of $F_X = 5.1^{+1.4}_{-0.9} \times 10^{-13}$ erg s$^{-1}$ cm$^{-2}$. Due to the low counts, the existing data cannot distinguish between a power law and an \texttt{apec} model, which produced an equally good fit (c-stat/d.o.f. = 0.99, vs c-stat/d.o.f. = 1.03 for the power law), with $kT=3.5^{+2.1}_{-1.0}$ keV. 

\begin{figure}[ht!]
\includegraphics[width=1.1\columnwidth]{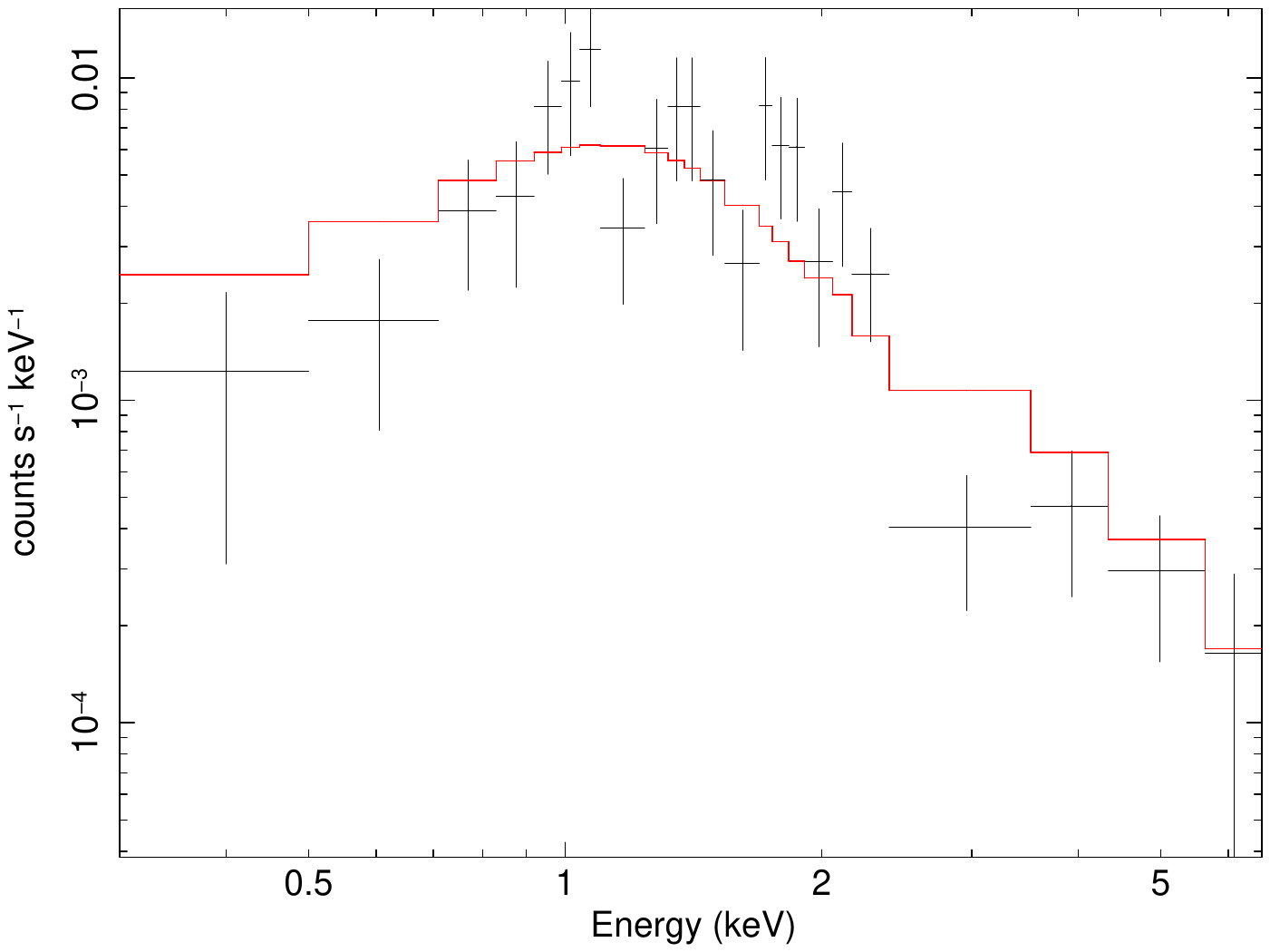}
\caption{Binned Swift spectrum of J1824A (black) and absorbed power law fit (red). The absorbed power law model is a good fit to the data (c-stat/d.o.f. = 110.43/107). \label{fig:j1824_swift_spec}}
\end{figure}

\subsubsection{Chandra Spectrum}

With $N_H$ free, the best-fit absorbed power law to the Chandra spectrum of J0640A (c-stat/d.o.f. = 398.56/441) gave a photon index $\Gamma = 1.93^{+0.33}_{-0.23}$ and $N_H = (6^{+19}_{-6}) \times 10^{20}$ cm$^{-2}$. With the absorbing column density fixed to the expected foreground value $N_H = 1.12\times10^{21}$ cm$^{-2}$, the best-fit model (c-stat/d.o.f. = 398.75/442) gave $\Gamma = 2.00^{+0.18}_{-0.17}$. This results in an unabsorbed 1.0---10.0 keV flux of $F_X = (3.8^{+0.5}_{-0.4})\times 10^{-13}$ erg s$^{-1}$ cm$^{-2}$. The flux is consistent with the XMM and Swift observations, but the spectral fit suggests a slightly softer power law at this epoch.

\subsubsection{eROSITA Spectrum}

We fit the J0640A eROSITA 0.2---2.3 keV spectrum with an absorbed power law and obtained a good fit with high uncertainty (c-stat/d.o.f. = 234.45/284) of photon index $\Gamma = 3^{+2}_{-1}$ and $N_H = 3^{+4}_{-2} \times 10^{21}$ cm$^{-2}$. With $N_H$ fixed to the expected foreground absorption, the fit was improved to $\Gamma = 1.9 \pm 0.4$, yielding a 0.2---2.3 keV flux of $F_X = (4 \pm 1) \times 10^{-13}$ erg s$^{-1}$ cm$^{-2}$. If we assume the best spectral model fit to the XMM spectrum with fixed $N_H$ and $\Gamma = 1.8$, the 1.0---10.0 keV unabsorbed flux in this observation is $(5\pm1)\times10^{-13}$ erg s$^{-1}$ cm$^{-2}$, which is consistent with the 1eRASS hard catalog flux of $(4.0 \pm 0.9) \times 10^{-13}$ erg s$^{-1}$ cm$^{-2}$ and with the flux measured at other epochs by XMM, Swift and Chandra.

\subsubsection{Summary of X-ray Spectral Results}

Although the distances to both sources are unknown, the X-ray spectra offer clues to the nature of the accreting compact objects. The lower bound on the 90\% confidence interval for the value of the photon indices are $\Gamma > 1.5$ and $\Gamma > 1.6$ respectively. These are softer X-ray spectra than typical for the intermediate polar (also known as DQ Her) sub-class of cataclysmic variables (CVs), which show hard X-ray spectra of $\Gamma = 1.0-1.3$ \citep{DQHer_Patterson1994, CV_Xrays_Mukai2017}. We also find no evidence for the 6--7 keV Fe emission line complex, which is commonly observed in the X-ray spectra of intermediate polars. Novalike variables, which are non-magnetic CVs in a persistently high accretion state, display X-ray spectra best described by the \texttt{apec} model. We have shown that an absorbed \texttt{apec} model is a worse fit to the spectrum of J0640A than a power law, and it is therefor unlikely that J0640A system hosts a white dwarf. More X-ray data are needed to make such a determination for J1824A based on X-ray data alone.

The fits to the mean X-ray spectra are, however, consistent with the spectra seen in tMSPs. PSR J1023+0038 has been observed to display a power-law spectrum of photon index $\Gamma \sim 1.7$, with little variation between disk-state X-ray luminosity modes \citep[e.g.,][]{CotiZelati2014, J1023_VLA_Chandra, J1023_Baglio2023}. Together with candidate tMSPs put forth in the literature, the observed X-ray spectra of disk-state tMSPs show power law shapes defined by a narrow range of photon index, from $\Gamma = 1.4$ in the case of the edge-on candidate 3FGL J0427.9–6704 \citep{J0427_Li2020} to $\Gamma = 1.8$ seen for 4FGL J0540.0-7552 \citep{J0540_discovery}. The X-ray spectra of J0640A and J1824A indicate that they are more consistent with disk-state tMSPs than accreting white dwarfs.

\subsection{X-ray Variability}\label{sec:xray_var}

\begin{figure*}[t!]
\centering
\includegraphics[width=\textwidth]{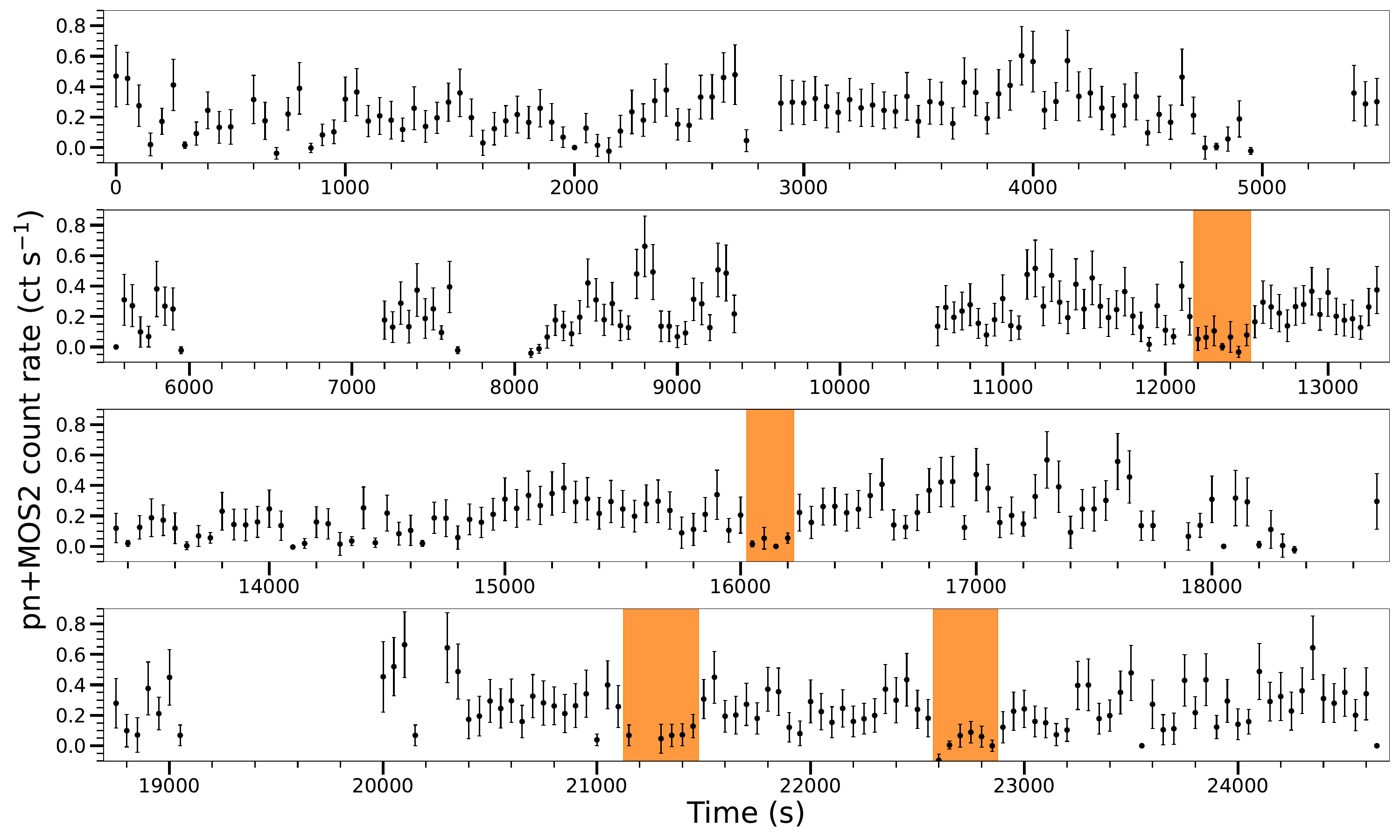} 
\caption{XMM light curve of J0640A binned in 50 s bins. The count rate varies by factors of 2--3 over a range of timescales, persistent around the mean count rate of 0.22 ct s$^{-1}$. Potential low modes are indicated by the orange shaded regions. Gaps with no data are durations of high background flaring.  \label{fig:xmm_lc}}
\end{figure*}

\begin{figure}[t!]
\includegraphics[width=1.0\columnwidth]{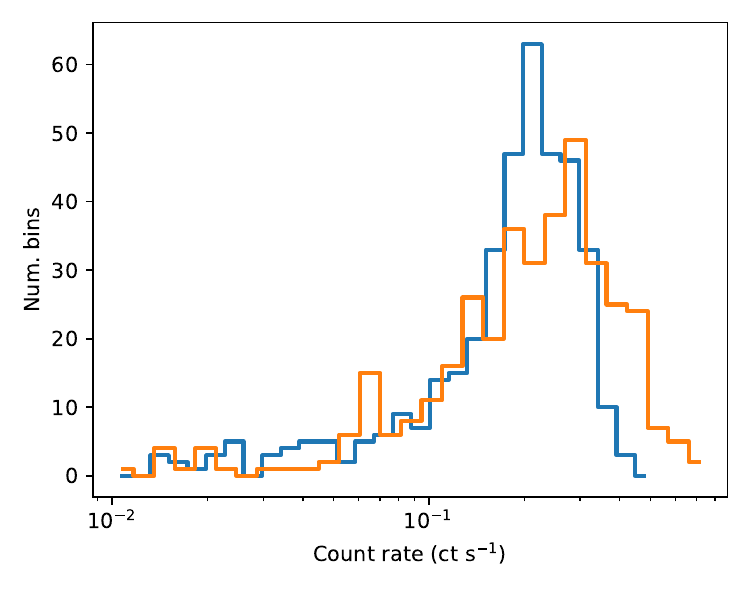} 
\caption{Orange: count rate distribution from the 50 s-binned XMM light curve in Figure \ref{fig:xmm_lc}. Blue: count rate distribution of a representative simulated light curve of PSR J1023+0038 scaled to the flux of J0640A. The distributions are from light curves with the same number of data points. The low mode peak in the original data of PSR J1023+0038 is not detected under the conditions of the J0640A observation. \label{fig:xmm_lc_hist}}
\end{figure}

There are two distinct forms of X-ray phenomenology observed in confirmed tMSP light curves that can distinguish strong candidate systems. The first is X-ray moding, where the X-ray count rate shows bimodal behavior and changes by a factor of $\sim 6$--7 on timescales of tens of seconds \citep[e.g.,][]{deMartino2013, Campana_diSalvo2018, J1023_Baglio2023}. The low mode is less common than the high mode, with tMSPs typically spending 20--30\% of their time in the low mode. This X-ray moding behavior has been observed in all three confirmed tMSPs and in the candidates 3FGL J1544.6–1125 \citep{Bogdanov2016} and CXOU J110926.4–650224 \citep{J1109_CotiZelati2021}, and is unique to this source class. It remains unclear whether disk-state tMSPs might experience periods where X-ray moding is inactive outside of flaring episodes. The non-flaring candidate tMSP 4FGL J0407.7--5702 has not shown clear evidence for moding \citep{J0407_discovery}, but it is possible that this is because it is more distant with lower count rates, and more sensitive future observations might well reveal X-ray moding in this system as well.

For J0640A, we show the XMM light curve with 50 s bins in Figure \ref{fig:xmm_lc} and the distribution of these count rates in Figure \ref{fig:xmm_lc_hist}. The light curve shows moderate, factor of $\sim 2$--3 fluctuations in count rate over a range of timescales, both between individual 50 s bins and also on longer timescales of tens of minutes. If X-ray moding was active at the time of this observation, then the 0.22 ct s$^{-1}$ mean count rate would represent the high mode of the system. The low state, if present, would peak about an order of magnitude fainter and last tens to hundreds of seconds. 

To determine if this data set is sensitive to tMSP low modes, we scaled an XMM observation of PSR J1023+0038 that showed clear X-ray moding \citep{J1023_Baglio2023} to the flux observed for J0640A. We binned the original light curves by 50 s and considered only the pn and MOS2 detectors. We scaled the count rates by the ratio between the mean count rate of the high mode of PSR J1023+0038 and the mean count rate of J0640, separately for each instrument to mimic the effect of the source being on a chip gap of the pn camera. For each 50 s time bin we resampled the counts from a Poisson distribution whose mean was the scaled number of counts in the bin. We then combined the pn+MOS2 light curves with lcmath in HEASOFT and applied a window function with a randomly sampled start time, corresponding to the duration and gaps in the observation of J0640A due to background flaring (no durations of high background flaring were detected in the light curve of PSR J1023+0038). We followed this procedure to produce $10^4$ simulated light curves of PSR J1023+0038 under the conditions of the J0640A observation. 

In the simulated light curves, long-duration low X-ray modes lasting $\gtrsim 200$ s are still identifiable. With the window function of the J0640A observation, five or less of these long-duration low modes are present in a given simulated light curve. The count rate distributions of the simulated light curves are dominated by the high mode, and a peak corresponding to the X-ray low mode cannot be identified. We show one of the simulated count rate distributions overlaid with the observed distribution for J0640A in Figure \ref{fig:xmm_lc_hist}. We find that the observation of J0640A is not sensitive to X-ray modes lasting less than 200 s, and a handful of longer low modes should be detected if the system undergoes moding similar to PSR J1023+0038. In Figure \ref{fig:xmm_lc}, we indicate four durations in the light curve of J0640A that could be low modes lasting $\sim$200--350 s. Longer low modes are ruled out in this source.

From two-sample Kolmogorov–Smirnov tests between the count rate distributions of each simulated PSR J1023+0038 light curve and the observed J0640A light curve, only 0.45\% of the simulated light curves resulted in a p-value $>0.05$. This indicates that J0640A is not drawn from the same distribution as the scaled PSR J1023+0038 light curve. Differences in the underlying distributions may arise from moding at a different ratio of high to low mode mean count rate, or simply from higher variability in the J0640A light curve than in the high mode of PSR J1023+0038. Given the broader distribution of J0640A seen in Figure \ref{fig:xmm_lc_hist}, the latter appears likely.

Another X-ray phenomenology observed in tMSPs is flaring, where the count rate increases by a factor of 4--8 over the mean rate \citep[e.g.,][]{deMartino2013}. At least two candidate tMSPs \citep{J0540_discovery, J0427_Li2020} appear to be dominated by frequent X-ray flares and have not been observed to show high and low modes, while other sources show moding and less frequent ($<10$ d$^{-1}$) flares (e.g., PSR J1023+0038, \cite{CotiZelati2014}; XSS J12270-4859, \cite{deMartino2013}; CXOU J110926.4–650224, \cite{J1109_CotiZelati2021}). The candidate 3FGL J1544.6–1125 has been observed at several epochs to switch between the high and low modes, but no flares have yet been observed in that system \citep{J1544_discovery, Bogdanov2016, Gusinskaia2024}, nor were flares observed in a 22 ksec observation of the candidate 4FGL J0407.7–5702 \citep{J0407_discovery}. X-ray flares are not detected in this observation of J0640A, but additional observations are needed to determine whether X-ray flares are absent over timescales longer than the $\sim6$ hr baseline of the XMM data, as for 3FGL J1544.6–1125. 

We performed a $\chi^2$ test for variability of the J1824A Swift light curve. We found evidence of variability with all time bin lengths used (0.5, 1.0, 1.5, 2.0 ksec), but the amplitude and timescale cannot be well characterized from the limited existing data. A long-baseline targeted observation of J1824A with XMM is needed to detect or rule out X-ray moding and flaring in this faint source.

\section{Optical Results}\label{sec:results:optical}
\subsection{Optical Spectroscopy}
\begin{figure*}[ht!]
\includegraphics[width=1.0\textwidth]{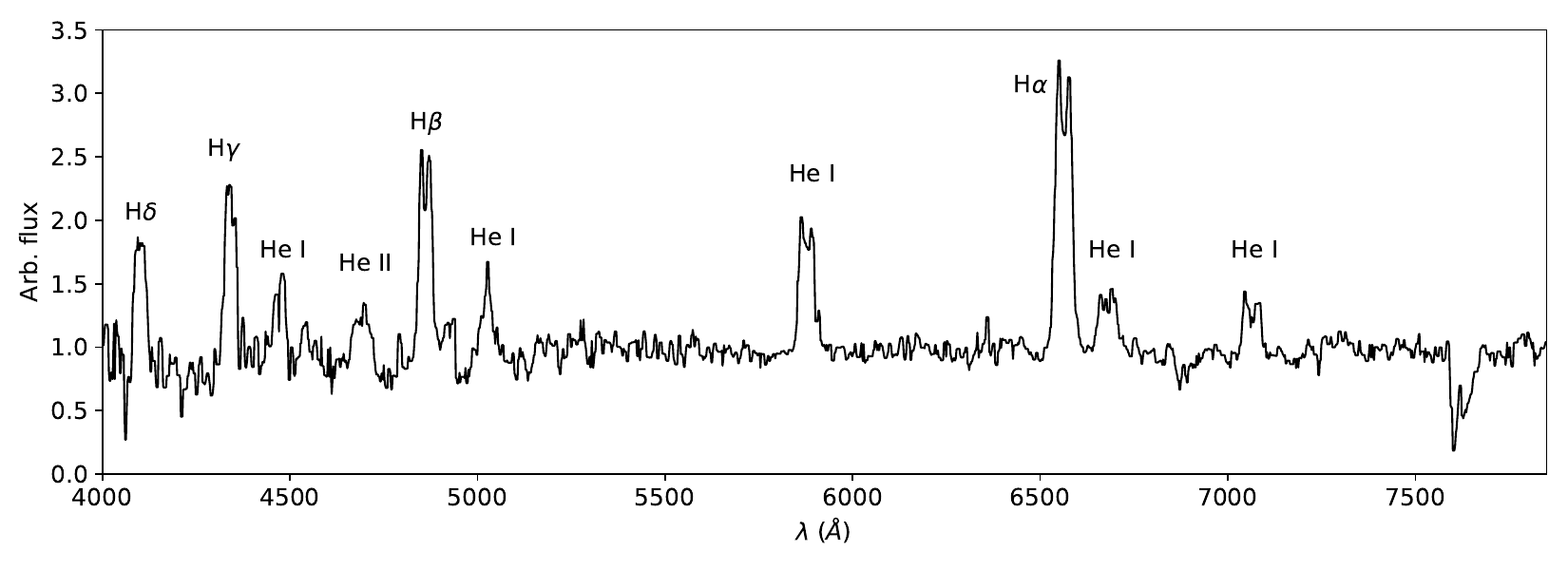}
\caption{Continuum-normalized and median-smoothed SOAR/Goodman spectrum of J0640A observed 2024 Mar 10. Double-peaked H and He emission lines characteristic of a hot accretion disk are clearly present, and the only absorption lines present in the spectrum are telluric. \label{fig:smooth_spec}}
\end{figure*}

\begin{figure*}[ht!]
\includegraphics[width=1.0\textwidth]{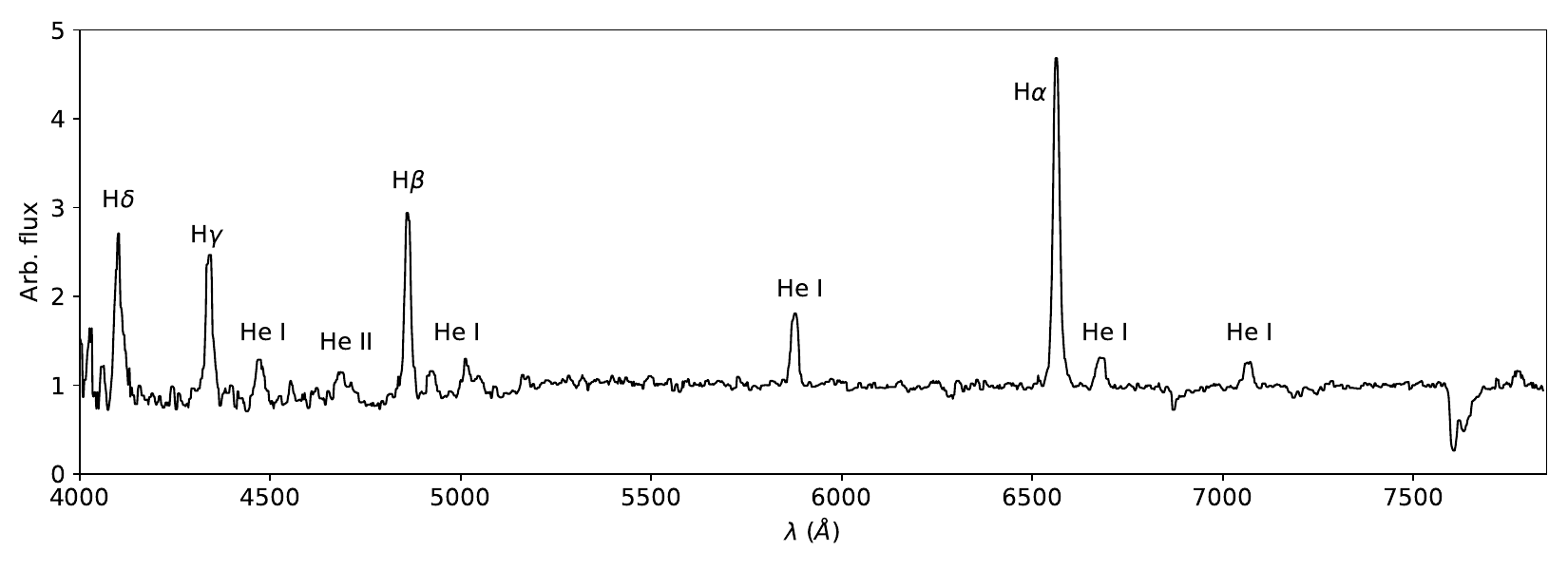}
\caption{Continuum-normalized and median-smoothed SOAR/Goodman spectrum of J1824A observed 2024 August 10. Strong H and He emission lines indicative of an accretion disk are clearly present, and the only absorption lines present in the spectrum are telluric. \label{fig:j1824_smooth_spec}}
\end{figure*}

We show representative smoothed spectra of J0640A and J1824A in Figures \ref{fig:smooth_spec} and \ref{fig:j1824_smooth_spec}. All 21 low-resolution optical spectra of J0640A show a flat continuum with broad double-peaked H and He emission lines, and an absence of absorption lines.  The mean S/N of the four medium-resolution spectra was low due to the faintness of the source (S/N $\sim$ 2--3), so we only consider measurements of the H$\alpha$ line from these spectra. All 17 spectra of J1824A show strong single-peaked H$\alpha$, H$\beta$ and H$\gamma$ lines in emission, as well as fainter He I emission lines at $\lambda 5015, 5875, 6678$ and 7065 \AA. In the higher S/N spectra obtained 2024 August 10, the helium profiles appear flat-topped and the H$\alpha$ profile in one spectrum shows similar structure, perhaps from double-peak broadening that is not quite resolved. H$\delta$ and additional helium lines at HeI$\lambda4471,4922,6678,7065 \AA$ and HeII$\lambda4686 \AA$ are also visible. The only absorption lines detected in spectra of both sources are telluric. 

We used the specutils \citep{specutils} and astropy \citep{astropy:2013, astropy:2018, astropy:2022} Python packages to measure the equivalent widths (EWs) and fit the emission profiles of all measurable Balmer and helium emission lines with a single-Gaussian model. We also fit the double-peaked profiles of the J0640A spectra with a two-Gaussian model to characterize the separation of the red- and blue-shifted peaks. We give the mean EW and single-Gaussian FWHM for each source, and the double-Gaussian peak separation for J0640A in Tables \ref{tab:mean_emission} and \ref{tab:j1824_mean_emission}. The standard deviations reported represent both systematic uncertainties and variability intrinsic to the source. The latter clearly dominates the standard deviation of the strongest lines which are detected in all spectra, including H$\alpha$, H$\beta$ and the 5875 \AA He I line. 

\begin{figure}[ht!]
\includegraphics[width=1.0\columnwidth]{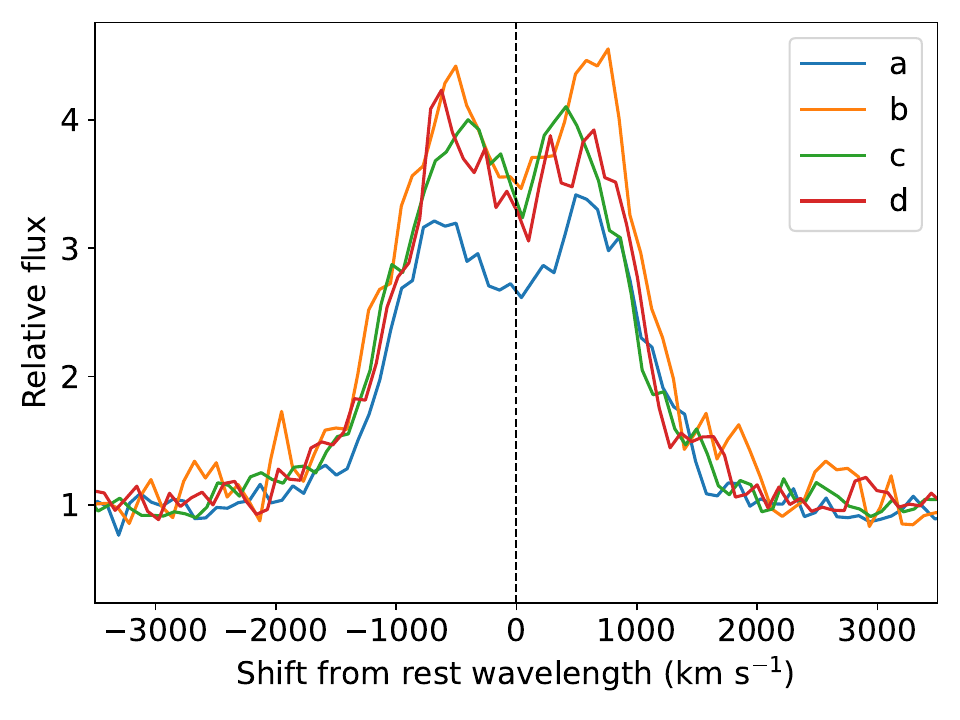}
\caption{Line profiles of H$\alpha$ from the SOAR/Goodman spectra of J0640A obtained on 2024 March 10. The observation times relative to the first spectrum and line measurements for each profile are given in Table 
\ref{tab:halpha_profiles}. Although similar heights, the maximum change is seen between profiles b (orange) and c (green), where the FWHM decreases by 11\% and the separation of the two peaks decreases by 10\% between these two 25-minute exposures separated by 114 minutes.  \label{fig:halpha_profiles}}
\end{figure}

The mean resolution-corrected FWHM of a single Gaussian fit to the H$\alpha$ profile across all obtained spectra of J0640A is 2171 km s$^{-1}$ with a standard deviation of 104 km s$^{-1}$. With two Gaussians fit to the line profile, the mean separation of the peaks of the H$\alpha$ profile is 1199 km s$^{-1}$ with a standard deviation of 107 km s$^{-1}$. The line profile is highly variable across each observation, including between spectra obtained within the same night as shown in Figure \ref{fig:halpha_profiles}. We also found strong variations in the H$\alpha$ radial velocities, which (among spectra with median S/N $> 4$) had a mean value of 147 km s$^{-1}$ and standard deviation 265 km s$^{-1}$. Within each observed spectrum there is a positive correlation between the transition potential energy and FWHM, such as H$\alpha$ vs. H$\beta$ and the He I 5875 \AA\ line vs. the He II 4685 \AA\ line, as expected for the model of an accretion disk where the fast-rotating inner disk is hotter than the outer regions.

\begin{figure}[ht!]
\includegraphics[width=1.0\columnwidth]{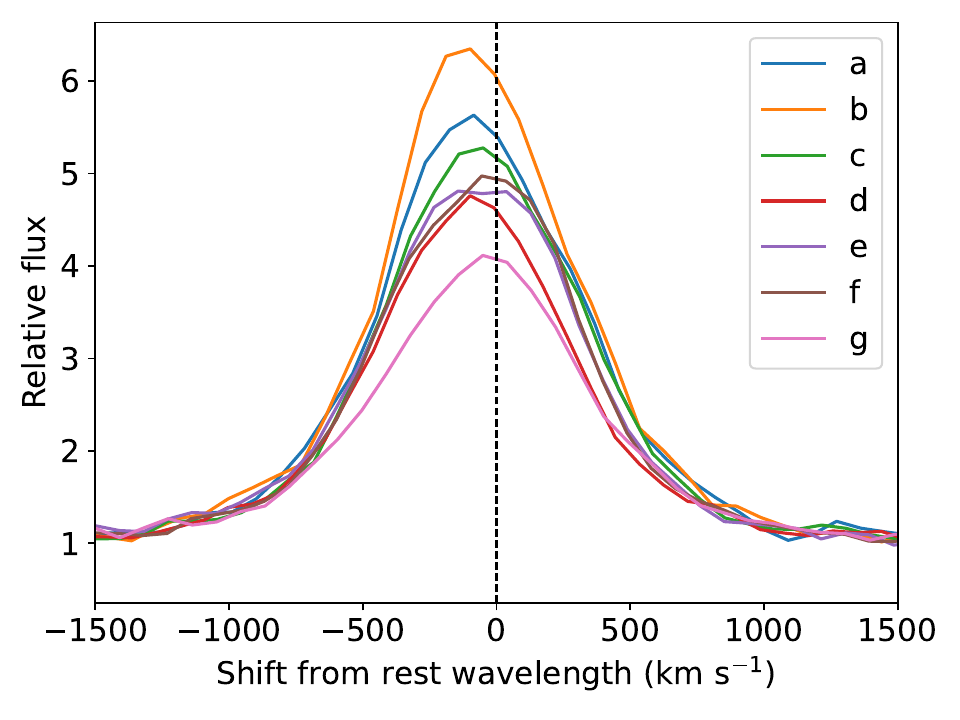}
\caption{Line profiles of H$\alpha$ from the SOAR/Goodman spectra of J1824A obtained on 2024 August 10. The observation times relative to the first spectrum and line measurements for each profile are given in Table 
\ref{tab:j1824_halpha_profiles}.  \label{fig:j1824_halpha_profiles}}
\end{figure}

The mean resolution-corrected FWHM of the H$\alpha$ emission line profile of J1824A is 957 km s$^{-1}$ with a standard deviation of 136 km s$^{-1}$ across the 17 spectra. Similar to the spectra of J0640A, we find a trend of FWHM increasing with the transition energy associated with each Balmer line. The mean RV of the H$\alpha$ line, considering only spectra with median $S/N > 10$ is --55 km s$^{-1}$ with a standard deviation of 20 km s$^{-1}$. In Figure \ref{fig:j1824_halpha_profiles} we show the H$\alpha$ line profile and how it varies in the highest S/N spectra, all obtained on 2024 August 10. Table \ref{tab:j1824_halpha_profiles} contains the profile measurements from each spectrum. The H$\alpha$ RV varies from --49 to --89 km s$^{-1}$ across about 90 minutes between integration time midpoints. 

Optical spectroscopy has shown that the emission features of tMSPs are broad, occasionally double peaked, and variable on multiple timescales. Measurements of the mean H$\alpha$ FWHM of tMSPs are typically $\gtrsim 1000$ km s$^{-1}$ \cite[e.g.,][]{Linares2014ATel, J1109:Coti_Zelati:2019, J0540_discovery}. In some systems, like PSR J1023+0038, the line morphology is persistently double-peaked with variable peak separations, while in other tMSPs the structure is unresolved in spectra at similar resolving power. Spectroscopic observations of all tMSPs have shown variability on the minimum timescales they are sensitive to, typically tens of minutes but also as short as 50 seconds in the case of high-cadence spectroscopy of PSR J1023+0038 by \cite{J1023:Messa:2024}. The flare-mode candidate tMSP 4FGL J0540.0–7552 shows the strongest emission features in the literature to date with mean H$\alpha$ FWHM $\sim 1910$ km s$^{-1}$ \citep{J0540_discovery}. The spectral features of J0640A presented here would be the strongest for any known tMSP or candidate, at least partially related to a likely high inclination with respect to our line of sight. The features of J1824A are more typical for a tMSP, lying within the range of values measured for other systems. We further discuss the optical spectral line measurements for these two sources in the context of tMSPs in Section \ref{sec:discussion:optical}.

\begin{deluxetable}{llcll}
\tablecaption{Mean and standard deviation of measured EW, FWHM for the single Gaussian model fit to the emission line profiles, and peak separation of the two-Gaussian fit for the H and He emission lines present in the optical SOAR/Goodman spectra of J0640A. \label{tab:mean_emission}}
\tablehead{\colhead{Line} & \colhead{Rest $\lambda$} & \colhead{EW} & \colhead{FWHM} & \colhead{Peak sep.} \\
           \colhead{}     & \colhead{\AA}   & \colhead{--\AA} & \colhead{km s$^{-1}$} & \colhead{km s$^{-1}$}}
\startdata
H$\delta$ & $4101.7$ & $21.2 \pm 6.5$ & $1894 \pm 749$ & $1409 \pm 437$ \\
H$\gamma$ & $4340.5$ & $42.3 \pm 8.3$ & $2288 \pm 294$ & $1412 \pm 273$ \\
He I & $4471.0$ & $15.1 \pm 3.5$ & $2001 \pm 638$ & $1335 \pm 507$ \\
He II & $4685.0$ & $30.1 \pm 11.9$ & $3204 \pm 702$ & $1649 \pm 774$ \\
H$\beta$ & $4861.4$ & $75.0 \pm 22.9$ & $2480 \pm 284$ & $1337 \pm 134$ \\
He I & $5015.0$ & $14.6 \pm 5.8$ & $1685 \pm 691$ & $1195 \pm 568$ \\
He I & $5875.0$ & $53.0 \pm 12.6$ & $2378 \pm 170$ & $1173 \pm 268$ \\
H$\alpha$ & $6562.8$ & $111.2 \pm 27.4$ & $2171 \pm 104$ & $1199 \pm 107$ \\
He I & $6678.0$ & $18.9 \pm 10.3$ & $2052 \pm 361$ & $1332 \pm 153$ \\
He I & $7065.0$ & $29.4 \pm 9.1$ & $2534 \pm 412$ & $1092 \pm 362$
\enddata
\end{deluxetable}

\begin{deluxetable}{lllllll}
\tablecaption{H$\alpha$ profile measurements from spectra of J0640A obtained 2024 March 10. The Time column gives the midpoint of the integration time in minutes since the first spectrum of the night. The S/N column is the median ratio of source flux to uncertainty per pixel} for the entire spectrum. The FWHM and RV are derived from the fit of a single Gaussian to the line profile. \label{tab:halpha_profiles}
\tablehead{\colhead{Profile} & \colhead{Time} & \colhead{S/N} & \colhead{EW} & \colhead{FWHM} & \colhead{RV} & \colhead{Peak sep.} \\
           \colhead{} & \colhead{min.} & \colhead{}    & \colhead{--\AA} & \colhead{km s$^{-1}$} & \colhead{km s$^{-1}$} & \colhead{km s$^{-1}$}}
\startdata
a & 0.0 & 7.2 & 105.2 & 2107 & 37.3 & 1241 \\
b & 27.1 & 4.9 & 169.3 & 2200 & 22.0 & 1227 \\
c & 141.0 & 6.2 & 140.2 & 1963 & -45.9 & 1101 \\
d & 166.9 & 5.0 & 138.7 & 2060 & -26.1 & 1171
\enddata
\end{deluxetable}

\begin{deluxetable}{llcl}
\tablecaption{Mean and standard deviation of measured EW, FWHM for the single Gaussian model fit to the line profiles of the H and He emission lines present in the optical SOAR/Goodman spectra of J1824A. We note that the He I and H$\delta$ lines are not detected in all spectra due to variations in S/N, so the standard deviation of the measurements for these lines is more representative of the measurement uncertainty. \label{tab:j1824_mean_emission}}
\tablehead{\colhead{Line} & \colhead{Rest $\lambda$} & \colhead{EW} & \colhead{FWHM} \\
           \colhead{}     & \colhead{\AA}          & \colhead{--\AA} & \colhead{km s$^{-1}$}}
\startdata
H$\delta$ & $4101.7$ & $23.8 \pm 2.6$ & $1161 \pm 136$ \\
H$\gamma$ & $4340.5$ & $26.3 \pm 10.3$ & $1389 \pm 296$ \\
He I & $4471.0$ & $10.2 \pm 2.7$ & $1148 \pm 138$ \\
He II & $4685.0$ & $10.4 \pm 2.0$ & $1923 \pm 314$ \\
H$\beta$ & $4861.4$ & $31.5 \pm 14.5$ & $1218 \pm 210$ \\
He I & $5015.0$ & $6.9 \pm 3.7$ & $1277 \pm 719$ \\
He I & $5875.0$ & $11.7 \pm 6.4$ & $980 \pm 309$ \\
H$\alpha$ & $6562.8$ & $53.5 \pm 27.6$ & $957 \pm 136$ \\
He I & $6678.0$ & $6.6 \pm 3.4$ & $1045 \pm 118$ \\
He I & $7065.0$ & $8.0 \pm 3.6$ & $1269 \pm 541$
\enddata
\end{deluxetable}

\begin{deluxetable}{llllll}
\tablecaption{H$\alpha$ profile measurements from spectra of J1824A obtained 2024 Aug 10. The Time column gives the midpoint of the integration time in minutes since the first spectrum of the night. The S/N column is the median ratio of source flux to uncertainty per pixel for the entire spectrum for the entire spectrum.  \label{tab:j1824_halpha_profiles}}
\tablehead{\colhead{Profile} & \colhead{Time} & \colhead{S/N} & \colhead{EW} & \colhead{FWHM} & \colhead{RV} \\
           \colhead{} & \colhead{min.} & \colhead{}    & \colhead{--\AA} & \colhead{km s$^{-1}$} & \colhead{km s$^{-1}$}}
\startdata
a & 0.0 & 14.5 & 94.8 & 818 & -66.6 \\
b & 25.4 & 13.0 & 104.9 & 767 & -72.0 \\
c & 61.1 & 16.8 & 87.4 & 804 & -52.9 \\
d & 86.6 & 17.4 & 75.3 & 791 & -89.2 \\
e & 119.4 & 18.3 & 83.4 & 828 & -68.7 \\
f & 147.4 & 16.2 & 82.8 & 800 & -56.1 \\
g & 178.8 & 17.6 & 66.2 & 858 & -48.5
\enddata
\end{deluxetable}

\subsection{Photometric Variability}\label{sec:results:opt_var}

\begin{figure}[ht!]
\includegraphics[width=1.0\columnwidth]{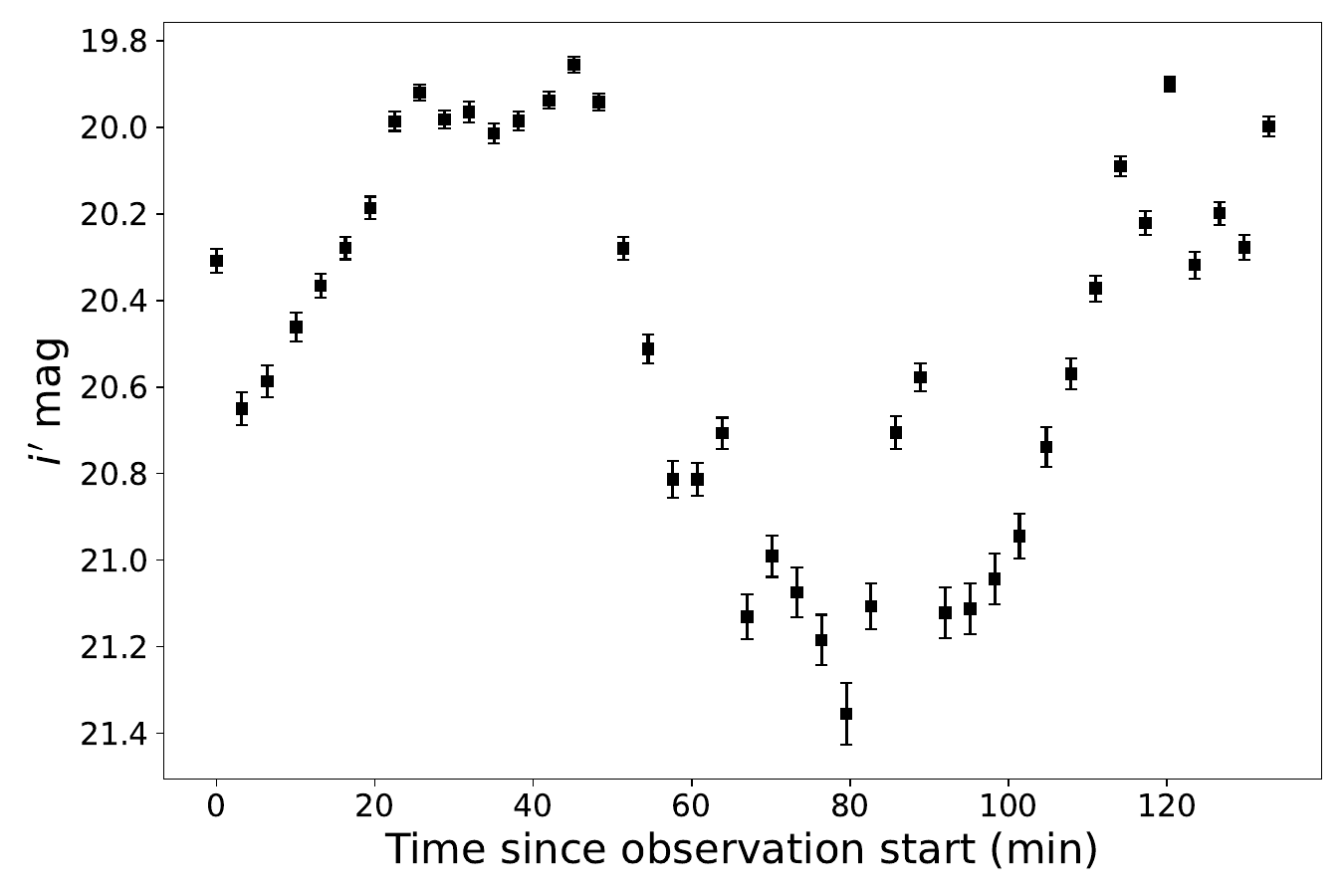}
\caption{SOAR/Goodman $i'$-band light curve of J0640A with a baseline of 2.2 hr, featuring a maximum change in brightness of 1.5 mag across 34 minutes. A longer baseline is necessary to show whether the light curve is periodic; attempts to fold the X-ray light curve and H$\alpha$ radial velocities on the apparent 102.2 min period did not result in periodicity of either property. Aperiodic optical variability of this amplitude and timescale is seen only in other disk-state tMSPs. \label{fig:soar_lc}}
\end{figure}

The optical light curve of J0640A shown in Figure \ref{fig:soar_lc} displays variability over $\sim1.5$ mag on the timescale of tens of minutes as well as $\sim 0.2-0.5$ mag amplitude variability over shorter durations. A Lomb-Scargle periodogram of the SOAR light curve yielded one significant peak in the power spectrum at 102.2 min, which visually matches the overall sinusoid-like shape of the variability. However, attempts to fold the XMM light curve and H$\alpha$ RV measurements on this period did not result in periodic behavior, hence there is no clear evidence that this variability arises from orbital motion. We also attempted separate period searches with the H$\alpha$ RV measurements and XMM light curve data, but did not find any significant peaks.

Optical ``flickering" is a commonly observed phenomenon in accreting compact objects, with disk-state tMSPs showing higher amplitudes of variability compared to cataclysmic variables with disks. The short-timescale variability in the light curve, including the $\sim 0.5$ mag flare at $\sim 86$ min, could alternatively be linked to the X-ray flaring state. The timescale and amplitude of the optical variability of J0640A is similar to those seen in observations of PSR J1023+0038 during durations of flaring \citep{Kennedy2018} and to the flare-mode candidate tMSP 4FGL J0540.0–7552, which showed aperiodic variability of amplitude $\sim1.8$ mag in $G$-band SOAR/Goodman photometry and in an $I$-band OGLE light curve \citep{J0540_discovery}. Although we did not detect flaring in the 2016 XMM light curve of J0640A, a new period of flaring could be active in the source in 2024 at the time of the optical observations. We interpret the 102.2 minute ``period" as being associated not with orbital motion but with variability of the accretion disk, whether through flickering or reprocessed X-ray flaring.

\begin{figure*}[ht!]
\includegraphics[width=1.0\textwidth]{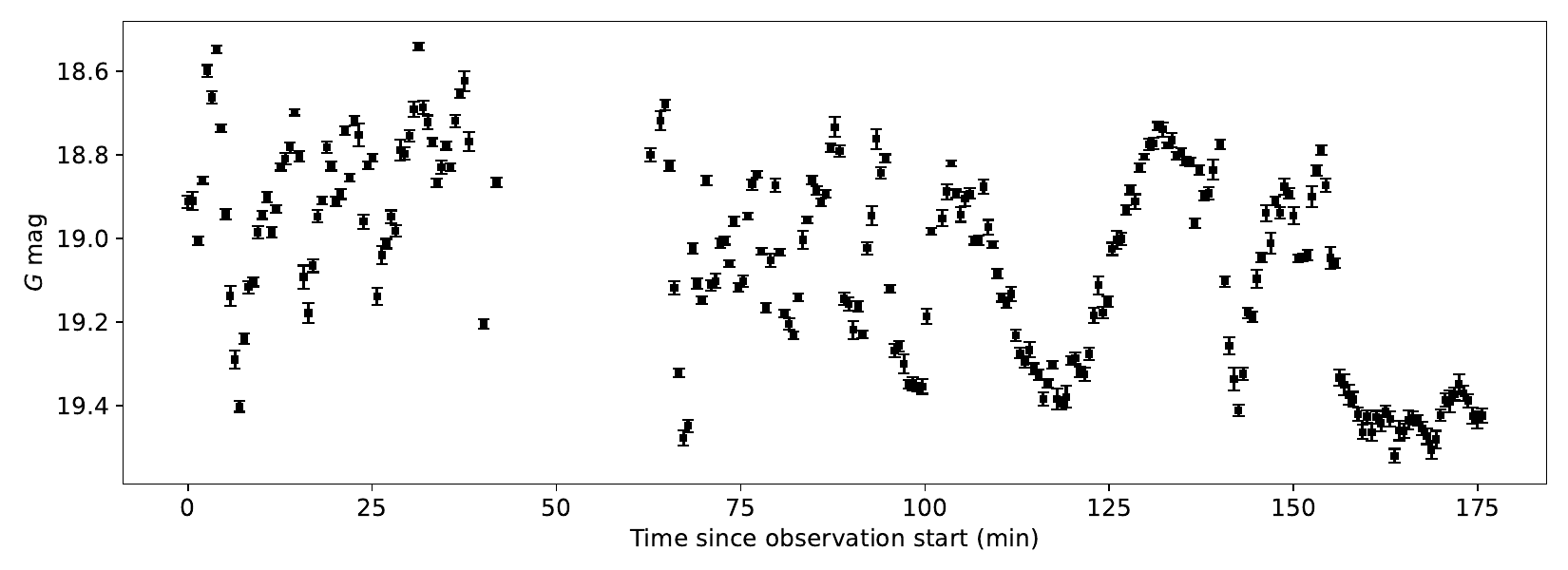}
\caption{SOAR/Goodman $G$-band light curve of J1824A with a baseline of 2.9 hr, featuring rapid limit-cycle behavior between $G\sim 18.7$--19.4. \label{fig:j1824_soar_lc}}
\end{figure*}

The SOAR/Goodman light curve of J1824A shown in Figure \ref{fig:j1824_soar_lc} displays strong variability between $G\sim 18.7$--19.4 over multiple timescales. A Lomb-Scargle periodogram of this light curve reveals one peak unassociated with the sampling cadence above the 0.01\% false alarm probability at 56.63 minutes. Two other peaks at the 1--5\% false alarm level are found at 24.7 and 103.3 minutes, and one much less significant peak at the 60\% false alarm level at 14.8 minutes. The shortest two timescales can be identified with the oscillations between $\sim18.7$--19.4 mag seen in the light curve, which are reminiscent of limit-cycle oscillations arising from the thermal-viscous instability produced by irradiation of the accretion disk. Toward the end of the observation window, the mean amplitude drops to about 19.5 mag and oscillates around this value with a much smaller amplitude than earlier in the observation. This could mark a temporary transition in the thermal structure of the accretion disk. The 103.3 minute timescale seems to be be produced by the gap in the observations around 50 minutes due to moving the telescope, which spans until the drop in mean brightness at the end of the observation. The light curve, with both consecutive sharp oscillations and slower smooth variations, bears a striking resemblance to high-cadence optical $V$-band photometry of the tMSP XSS J12270-4859 when it was in the disk state \citep{J1227:Pretorius:2009, J1227:deMartino:2010}.

The PS1 and ZTF light curves of J1824A indicate that its present optical brightness has been consistent over at least 13 years. The strong variability observed in the short-term light curve is also seen in the archival observations, with much sparser time sampling that shows maximum brightness changes of 1.1, 0.4, 0.6, 0.5 and 0.8 mag in the Pan-STARRS \textit{grizy} filters respectively. In the ZTF $g$ and $r$ filters observed between 2018--2022, the maximum brightness changes are 1.2 and 1.4 mag respectively. Separate Lomb-Scargle periodograms of the two archival light curves by filter produced two peaks in the ZTF $r$ filter power spectra just above the 1\% false alarm level, corresponding to 15.67 d and 15.95 d. A multi-band power spectrum, as implemented by astropy's LombScargleMultiband, also produced a peak close to the peaks in the periodogram of the $r$-band light curve at 15.92 d. This period does not appear to be related to the observing cadence of ZTF. No significant peaks were present in the power spectra of the Pan-STARRS light curves, by individual band nor multiband. Phase-folding the archival light curves on each of these periods did not reveal periodicity. Without a strong signal in both periodograms, it appears likely that the two long-term light curves sample the rapid oscillations seen in the SOAR/Goodman light curve of the source. 

\section{Discussion} \label{sec:discussion}

\subsection{Inspection of Additional X-ray Sources Within 4FGL J0639.1-8009 \label{sec:discussion:agn}} 

The brightest X-ray source in the 68\% positional uncertainty of 4FGL J0639.1-8009 is detected by five different X-ray observatories as described in Section \ref{sec:data}. Optical follow-up described in the following sections shows that the X-ray source is a binary system hosting a hot accretion disk around a compact object of uncertain nature. Of the possible compact binaries that could produce these signatures, only tMSPs are persistent $\gamma$-ray sources. Due to the large ($7.6\arcmin \times 6.3\arcmin$) 68\% positional uncertainty of this Fermi source, we inspected available X-ray data for additional sources that could persistently emit at $\gamma$-ray energies. 

As mentioned in Section \ref{sec:data:xray}, the AGN candidate 4XMM J063720.2-801230 is the second brightest source in the 68\% positional uncertainty of 4FGL J0639.1-8009, with an observed X-ray flux at least four times fainter in all catalogs than the compact binary we present as the more likely counterpart to the $\gamma$-rays. As of 4FGL DR3, 98\% of Fermi AGN are blazars \citep{Ajello2022_fermi_agn}. These are AGN oriented with their jet directed toward the observer, and so appear as radio-loud sources with spectral energy distributions characterized by two broad components from synchrotron emission at low energies and various processes at $\gamma$-ray energies. As part of an effort to identify new blazar associations for Fermi sources, \cite{DAbsrusco2014_WISE_blazars} used a ratio of radio to infrared flux ($q_{22}$) to define ``radio loudness" for a cross match between WISE and various radio catalogs. They adopted a threshold of $q_{22} < -0.5$ to mark radio-loud sources, and showed that confirmed $\gamma$-ray emitting blazars are reliably identified by this cut. 

The AGN candidate under consideration has an IR counterpart in the AllWISE catalog under the designation J063719.67-801230.2, with 22 $\mu$m band magnitude $W4 = 7.819 \pm 0.116$ \citep{WISE, AllWISE}. We inspected RACS imaging of the field to check for a radio detection, and measured a $3\sigma$ upper limit of 0.6 mJy at the source's position. We find a radio to IR flux ratio as defined by \cite{DAbsrusco2014_WISE_blazars} of $q_{22} > 0.97$, which shows that this AGN is not radio loud, and is therefor unlikely to be a blazar. Of the non-blazar AGN detected by Fermi, 65\% are radio galaxies, which are similarly ruled out by the source's radio-faintness. Due to the large angular size of Fermi-LAT positional uncertainties and the isotropic ubiquity of background galaxies, chance coincidence between unrelated galaxies and Fermi-LAT sources is highly likely. Indeed, the surface density of $W4 < 8$ mag AllWISE sources outside of 5 deg of the Galactic plane is about 122 deg$^{-2}$, so 5--6 coincident bright IR sources are expected in the 68\% positional uncertainty of 4FGL J0639.1-8009. 

As a radio-quiet AGN, several of which are expected to be located within the positional uncertainty of a given Fermi-LAT source, we believe that the candidate AGN 4XMM J063720.2-801230 is unlikely to be associated with 4FGL J0639.1-8009. The following multiwavelength results relate to the compact binary system corresponding to the brightest X-ray source in the Fermi ellipse.

\subsection{Could These Systems be Accreting White Dwarfs?}

For both systems presented here, the optical spectra and association with variable X-ray sources are clear evidence for compact binary systems with low-mass companions and accretion disks. While the coincidence with $\gamma$-ray sources is suggestive, given the several-arcminute angular size of the 68\% error ellipses for each source, a chance coincidence with a compact binary cannot be excluded a priori. We give the possibility of various classes of accreting white dwarfs due consideration. 

There are several accreting white dwarf configurations that produce variable X-ray emission and disk-dominated optical spectra that must be considered. For both systems we have presented, the lack of outbursts in the optical light curves rules out the possibility of dwarf novae in a standard cataclysmic variable. The measured X-ray photon indices are inconsistent with the hard X-ray spectra produced by intermediate polars \citep[IPs; ][]{Aizu1973}, a class of moderately magnetic accreting white dwarfs which was an early suggestion for the nature of PSR J1023+0038 on the basis of its variable radio emission \citep{Bond2002}. X-ray spectra of IPs exhibit strong Fe emission in the 6--7 keV band \citep[e.g.,][]{Mondal2024}, which we do not find evidence for in either system presented here. IPs exhibit power at multiple frequencies in their X-ray light curves attributed to orbital, spin, beat and side-band frequencies \citep{Norton1996}. We did not detect significant power at any frequencies in a Lomb-Scargle periodogram of the XMM light curve of J0640A. A long-baseline observation of J1824A with XMM is needed to perform similar analysis. In their optical spectra, IPs can be distinguished from non-magnetic CVs by their strong He II $\lambda4685 \AA$ EW ($\gtrsim 10 \AA$) and emission line ratio He II/H$\beta \gtrsim 0.5$ \citep[][and references therein]{Warner1995}. While the He II EW measured for J0640A and J1824A is high, the line ratio is lower than the IP regime (0.45 and 0.33 respectively). We find no evidence for either system to be an IP.

Non-magnetic white dwarfs with persistently high-accretion rates, known as novalikes, also must be considered. Novalikes have X-ray spectra best described by a thermal optically thin plasma model (\texttt{apec}) rather than the power low that is an excellent fit to tMSP spectra. We compared the fit of an absorbed single-temperature optically thin plasma model (\texttt{tbabs*apec}) to the best-fit power law models described in Section \ref{sec:results:xray:spectra}. For J0640A, we fit the XMM/pn+MOS2 spectrum with the absorbed \texttt{apec} model and could not obtain an acceptable fit even when we allowed the metal abundances to vary. The XMM spectrum of J0640A rules out the \texttt{apec} model for this system, making a novalike classification highly unlikely. For J1824A, we fit the Swift/XRT spectrum with an absorbed \texttt{apec} model with the $N_H$ fixed at the expected foreground value, and obtained a fit that is of similar quality to the power law fit. We therefore cannot rule out a novalike classification for J1824A based on the X-ray spectrum alone with the existing data.

Turning to their optical emission, novalikes are variable sources that display $\sim 0.2$--0.6 mag typical amplitudes of optical flickering on timescales of tens of minutes \citep{Bruch1992}. The VY Scl subclass of novalikes, which are defined by their optical high and low accretion states, can display stronger amplitudes of flickering when in their active state \citep{Fritz1998, Bruch2021}. The $\sim1.5$ mag and $\sim0.9$ mag amplitudes of optical variability seen in the respective SOAR/Goodman light curves of J0640A and J1824A are significantly higher than typical novalike optical flickering. The most extreme example of novalike flickering we could find in the literature was of the source MV Lyr, which was observed to fluctuate over an anomalously wide range of 1.4 mag in 16 minutes as the system transitioned to the high accretion state typical of VY Scl types \citep{Wenzel1989}. VY Scl types spend most of their time in an optical high state, with occasional low states that last weeks to a year and exhibit much narrower emission features \citep{book:warner:2003}. We observed persistent broad emission lines for both J0640A and J1824A before and after acquisition of their optical SOAR light curves, making it unlikely that either system is a VY Scl novalike observed during an optical state transition. The Pan-STARRS and ZTF light curves of J1824A also show that this system has been at the same mean brightness with consistent amplitudes of variability in each optical filter for over a decade.

\subsection{Comparison to Transitional Millisecond Pulsars}
tMSPs display a variety of highly variable multiwavelength signatures that allow candidate systems to be identified in the disk state before a transition has been observed. The following discussion of observed tMSP properties refers to their disk state only, as the properties differ substantially from the radio pulsar state. In Table \ref{tab:multi_comparison} we summarize relevant multiwavelength properties of the three confirmed tMSPs and candidate systems in the Galactic field.

\begin{deluxetable*}{lccccccc}
\tablewidth{1.0\textwidth}
\tablecaption{Comparison of multiwavelength properties in the sub-luminous disk state of confirmed tMSPs, previously published disk-state candidates, and the two new candidates presented here.} \label{tab:multi_comparison}
    
\tablehead{\colhead{Name} & \colhead{Distance\tablenotemark{a}} & \colhead{Orbital period\tablenotemark{b}} & \colhead{$\Gamma_X$\tablenotemark{c}} & \colhead{$F_X$\tablenotemark{d}} & \colhead{$F_X/F_{\gamma}$\tablenotemark{e}} & \colhead{$G$\tablenotemark{f}} & \colhead{$S_{\nu}$\tablenotemark{g}} \\
    \colhead{} & \colhead{kpc} & \colhead{h} & \colhead{} & \colhead{$10^{-13}$ erg s$^{-1}$ cm$^{-2}$} & \colhead{} & \colhead{mag} & \colhead{$\mu$Jy}
}
\startdata
\textit{Confirmed tMSPs} & & & & & & & \\
M28-I\tablenotemark{h} & 5.5 & 11.03 & $1.51 \pm 0.02$ & 6 & -- & -- & -- \\
PSR J1023+0038 & $1.37 \pm 0.04$ & 4.75 & $1.62 \pm 0.02$ & 80 & 0.22 & 16.2 & 122 (6 GHz) \\
XSS J12270--4859 & $2.0^{+0.7}_{-0.4}$ & 6.91 & $1.70 \pm 0.02$ & 152 & 0.33 & $V \sim 16.1$ & 353 (5 GHz) \\
\hline 
\textit{Candidate tMSPs} & & & & & & & \\
4FGL J0407.7--5702 & -- & -- & $1.74 \pm 0.04$ & 4 & 0.25 & 20.1 & -- \\
3FGL J0427.9--6704\tablenotemark{i} & $2.9^{+0.7}_{-0.5}$ & 8.80 & $1.44 \pm 0.05$ & 27 & 0.31 & 17.7 & 290 (5.5 GHz) \\
4FGL J0540.0--7552 & -- & -- & $1.78 \pm 0.07$ & 3 & 0.09 & 20.2 & $< 1500$ (5 GHz) \\
CXOU J110926.4--650224 & -- & -- & $1.63 \pm 0.02$ & 37 & -- & 20.1 & 33 (1.3 GHz) \\
3FGL J1544.6--1125 & $3.8 \pm 0.7$ & 5.80 & $1.73 \pm 0.04$ & 33 & 0.27 & 18.6 & 30 (6 GHz) \\
J0640A & -- & -- & $1.7 \pm 0.2$ & 5 & 0.23 & 20.6 & $<500$ (1.7 GHz) \\
J1824A & -- & -- & $1.9 \pm 0.3$ & 6 & 0.24 & 19.3 & $<200$ (0.9 GHz)
\enddata
\tablenotetext{a}{Distances with 68\% uncertainties, refined for XSS J12270--4859 and 3FGL J0427.9--6704 from zeropoint-corrected Gaia DR3 parallax measurements \citep{Lindgren2021}. They are in agreement with previous calculations based on DR2 parallaxes.}
\tablenotetext{b}{See \cite{J1109:Coti_Zelati:2019} and references therein.}
\tablenotetext{c}{Mean photon index for a simple absorbed power law model.}
\tablenotetext{d}{Mean unabsorbed X-ray flux in the 0.5--10 keV band for the purposes of calculating the X-ray to 0.1--100 GeV $\gamma$-ray flux ratio.}
\tablenotetext{e}{0.5--10 keV X-ray to 0.1-100 GeV $\gamma$-ray flux ratio. The $\gamma$-ray fluxes used are from 4FGL-DR4 (except XSS J12270--4859 which is taken from analysis of its pre-transition light curve by \cite{Torres2017}) and so previously reported flux ratios are updated here.}
\tablenotetext{f}{Gaia DR3 mean apparent magnitude, except in the case of XSS J12270--4859 which has been in the radio pulsar state for the duration of the Gaia mission; we instead list the $V$-band magnitude measured when it was in the disk state.}
\tablenotetext{g}{Mean flux density or $3\sigma$ upper limit at radio wavelengths, with the central frequency of each observation listed in parentheses \citep[for 4FGL J0540.0--7552, see][]{j0540:radio:schinzel2017}. Observations at $\sim10$ GHz are also available for some sources, but we only report flux densities at central frequencies as close as possible to those that the upper limits for J0640A and J1824A were observed with. The flat ($\alpha \sim 0$) slope of the radio spectrum allows for direct comparison between these quantities.}
\tablenotetext{h}{Due to its location in a globular cluster, the tMSP M28-I is not accessible with Fermi and Gaia. Its disk-state optical counterpart was identified with HST by \cite{M28I:hubble:pallanca2013}. We note that M28-I is the only known tMSP to undergo Type I X-ray bursts and show clear evidence that accreted material reaches the neutron star's surface. The only available radio flux density measurement was during an outbursting episode \citep{M28-I:xray_bursts:ferrigno2014}, and so we do not include it here since it is not directly comparable to the other sources. The source is currently in the radio pulsar state \citep[see][]{M28I:chandra_gbt:vurgun2022}.}
\tablenotetext{i}{All values for 3FGL J0427.9--6704 reference out-of-eclipse measurements.}
\tablecomments{See Table 6 of \cite{J1109:Coti_Zelati:2019} for additional tMSP properties related to multiwavelength emission, variability and binary configuration.}
\end{deluxetable*}

\subsubsection{X-ray Properties}

tMSPs display a distinct set of X-ray phenomenology. We have found that the mean X-ray spectra of J0640A and J1824A are consistent with the spectral shape observed for tMSPs, which are best fit by a power law spectral model with the narrow range of photon indices $\Gamma=1.6$--1.8 \citep{Linares2014, J0407_discovery}. As another distance-independent comparison, \cite{J0407_discovery} showed that disk-state tMSPs display high ratios of 0.5---10 keV X-ray flux to 0.1--100 GeV $\gamma$-ray flux in the range $F_X/F_{\gamma} = 0.26-0.43$. The discovery of candidate tMSP 4FGL J0540.0–7552 widened this range to $F_X/F_{\gamma} = 0.11-0.43$ \citep{J0540_discovery}. For J0640A, the flux ratio using the spectral fit to the mean XMM data with the MOS2 observation only (because the source lies on a chip gap of the pn camera) is $F_X/F_{\gamma} = 0.23^{+0.09}_{-0.06}$, with the uncertainty calculated using the 90\% confidence range of the X-ray flux. With the mean Swift/XRT data, which was observed over a longer time period starting six months after the XMM observation, the flux ratio is consistent at $F_X/F_{\gamma} = 0.24\pm0.03$. For J1824A, the flux ratio from the Swift/XRT observation is $F_X/F_{\gamma} = 0.24^{+0.05}_{-0.04}$. For both sources, the flux ratios are each well within the observed range seen for other tMSPs in the sub-luminous disk state.

tMSPs are highly variable X-ray sources, exhibiting bimodal X-ray emission states and strong flares. The observation of PSR J1023+0038 presented by \cite{J1023_Baglio2023} shows strong consecutive flares followed by X-ray moding without flares for at least 38 ksec, from the last small flare in the light curve to the end of the observation. Other tMSPs show an extreme range of X-ray flare duty cycles: candidate tMSPs include sources in a state of constant flaring \citep[e.g.,][]{J0427_Li2020, J0540_discovery} and sources which have not yet been observed to flare \citep[e.g., ][]{J1544_discovery, J0407_discovery}. For J1824A, variability is indicated in the existing Swift data, but a longer-baseline continuous observation by XMM is needed to investigate the presence of X-ray moding and flaring. The XMM light curve of J0640A presented here rules out flaring at the time of the observation for the $\sim25$ ksec baseline, but additional observations are needed to determine whether flares are absent entirely. We showed in Section \ref{sec:xray_var} that the existing XMM observation of J0640A is not sensitive to low modes lasting $< 200$ s, and indicated four potential low mode detections lasting $\sim$200--350 s while ruling out longer low modes in this source. While the evidence for moding in J0640A is marginal, the mean X-ray properties are consistent with those of tMSPs and the existing data do not rule out the possibility of X-ray moding in either source.

\subsubsection{Optical Properties}\label{sec:discussion:optical}

The optical spectra of tMSPs show a blue continuum dominated by line emission from the Balmer series and helium lines typical of an accretion disk. Measurements of PSR J1023+0038 by \cite{CotiZelati2014} showed persistent double-peaked emission lines, with mean H$\alpha$ EW $\sim 30$ \AA, FWHM $\sim1300$ km s$^{-1}$ and peaks separated by $\sim720$ km s$^{-1}$, and other observations have shown persistently strong features \citep[e.g.,]{J1023:Takata:2014, Linares2014ATel}. Recent high-cadence ($\sim50$ s time sampling) spectroscopy by \cite{J1023:Messa:2024} revealed erratic emission feature variability on timescales of minutes, including the occasional emergence of a third peak in the profile of H$\alpha$. These measurements, without any correlation between line flux or broadening to orbital motion or continuum flux, provided strong evidence for anisotropy in the accretion disk. The tMSP XSS J12270--4859 showed similar mean spectral features to PSR J1023+0038 during its disk state, although with the addition of Fe I emission and fainter lines overall leading to less structure clearly detected in the emission line profiles \citep{J1227:Masetti:2006, J1227:Pretorius:2009}.

The optical spectra of candidate tMSPs are consistent with the confirmed systems, with some variation in line morphology and the strength of absorption lines related to differences in inclination and the relative contributions of the accretion disk, intrabinary shock and companion. The two candidates with detected X-ray moding show H$\alpha$ EW $\sim 32$ \AA, with absorption features clearly detected in some spectra of 3FGL J1544.6--1125 and only weakly detected in stacked low-resolution spectra of CXOU J110926.4--650224 \citep{J1544_orbital, J1109:Coti_Zelati:2019}. The flare-mode candidate 4FGL J0540.0--7552 showed the strongest mean H$\alpha$ features of any tMSP in the literature until now, with EW $\sim 62$ \AA, FWHM $\sim1910$ km s$^{-1}$ and separation in the double-peaked profile of $\sim960$ km s$^{-1}$ \citep{J0540_discovery}. 

The optical spectra of J0640A and J1824A are dominated by broad Balmer and helium emission lines, and no absorption lines related to the companions can be identified even in the highest S/N spectra for each source. Much like other tMSPs, the line profiles and strengths for both sources are strongly variable on timescales of about one hour, the minimum that these data are sensitive to. All line profiles in the spectra of J0640A are double peaked, with mean peak separation in the H$\alpha$ profile of $\sim1199$ km s$^{-1}$. Despite being the faintest known tMSP in broadband photometry ($G=20.56$), J0640A shows the strongest mean H$\alpha$ emission of any tMSP with EW $\sim111$ \AA\ and FWHM $\sim2171$ km s$^{-1}$. The latter suggests a high inclination with respect to our line of sight and/or a relatively short orbital period. If the strong optical emission features of the candidate 4FGL J0540.0–7552 are related to its persistent X-ray flaring state, then the extreme emission features of J0640A may also support the idea that this source is in a new phase of active X-ray flaring that was absent in the previous XMM detection.

The lines in the spectra of J1824A are primarily single-peaked, although the profiles of the helium lines in some of the high S/N spectra appear to be flat-topped and there is structure in one epoch of the highest S/N spectra that suggests an unresolved separation of two peaks in H$\alpha$. The mean H$\alpha$ EW is $\sim54$ \AA, stronger than PSR J1023+0038 despite a likely distance much further away (see Section \ref{subsec:distances}). The mean H$\alpha$ FWMH of $\sim957$ km s$^{-1}$ is more intermediate for a tMSP, as expected from the much weaker detection of structure in the line profiles that suggests lower inclination than for J0640A.

tMSPs can display photometric variability related to reprocessed X-ray moding and flaring, aperiodic accretion disk instabilities, a compact jet, and orbital modulation. Simultaneous X-ray and optical observations of the prototypical tMSP and the candidate CXOU J110926.4--650224 show optical emission that echos the X-ray phenomenology at multiple epochs \citep[e.g.,][]{j1023:shahbaz:2015, Kennedy2018, J1023_Baglio2023, J1109:Coti_Zelati:2024}. Yet at other epochs, flares and bimodal switches are absent in optical observations \citep[e.g. PSR J1023+0038,][]{j1023:baglio:2019}. The cause of such a dramatic change in the efficiency of reprocessing X-ray emission to the optical regime is unclear. Simultaneous multiwavelength observations of PSR J1023+0038 by \cite{j1023:flares:baglio2025} alternatively suggest that flares in the X-ray and UV could be arise from the inner accretion disk becoming thicker and providing a larger shock region, while the optical and radio flares could arise from brief increases in the mass-loading efficiency of the jet as more free electrons become available. Observations of additional systems are necessary to explore whether this model is consistent with other tMSPs and constrain how such an increase of the disk thickness could occur.

As discussed in Section \ref{sec:results:opt_var}, the timescale and amplitude of the low-frequency optical variability of J0640A is similar to those seen in observations of PSR J1023+0038 during durations of flaring \citep{Kennedy2018} and to the amplitude $\sim1.8$ mag aperiodic variability of the flare-mode candidate tMSP 4FGL J0540.0–7552 \citep{J0540_discovery}. Although no X-ray flares were detected in the 2016 XMM observation of the source, the optical light curve suggests that a new episode of X-ray flaring could be active in J0640A. Indeed, the $\sim 0.5$ mag flare at $\sim 86$ min is consistent with the timescale of X-ray flares seen in other sources, and could potentially be a flare reprocessed from the X-ray to optical. In the light curve of J1824A we find no evidence for optical flaring. Instead, the light curve displays limit-cycle behavior over an amplitude of $\sim 0.7$ mag, closely resembling behavior seen at times in disk-state optical light curves of PSR J1023+0038 \citep{J1023:bogdanov:2015}, XSS J12270--4859 \citep{J1227:Pretorius:2009, J1227:deMartino:2010}, and the candidate 3FGL J1544.6--1125 \citep{J1544_discovery}. The photometric properties of both sources are compelling evidence for classification as candidate tMSPs.

tMSPs and other redbacks host pulsars in tight orbits with their stellar companions, with 68\% of observed orbital periods of Galactic field redbacks in the range 0.20--0.59 d \citep{rb_period:j1023:archibald2010, rb_period:j2339:romani2011, rb_period:j1816:kaplan2012, rb_period:j2129:ray2012, rb_period:j1723:crawford2013,Bassa2014,rb_period:j1626:li2014,rb_period:j0523:strader2014,rb_period:j1431:bates2015, rb_period:j2039:salvetti2015,rb_period:j2129:broderick2016,rb_period:j1048:deneva2016,rb_period:j0212:li2016, rb_period:j1957:stovall2016,J0427_Strader2016,J1544_orbital,rb_period:j0838:halpern2017,rb_period:j0954:li2018,rb_period:j1306:linares2018,rb_period:j1702:corbet2022, rb_period:j1910:au2023,rb_period:j1036:clark2023,rb_period:j1803:clark2023,rb_period:j2055:lewis2023,rb_period:j1622:turchetta2023,rb_period:j1646:zic2024, rb_period:j1908:simpson2025}. As discussed in Section \ref{sec:results:optical}, searches for orbital signatures in the optical and X-ray emission did not yield any candidate orbital periods, and we did not detect any absorption lines from the companion of either source. We also searched for absorption features in the co-added spectra, but none were revealed. Stacking the spectra without correcting for the unknown radial velocities associated with orbital motion at each epoch effectively smears out the spectral features, so weak absorption lines were not expected to be detected with this method. Spectroscopic observations with a larger telescope will be necessary to detect signatures from the companion, particularly with temporal sensitivity, although these signals may be lost within the variability of the much brighter accretion disk until the radio pulsar state emerges. 

\subsection{Distance Estimates} \label{subsec:distances}

As primarily faint sources, obtaining precise distances to tMSPs is challenging. The strongest distance constraints exist for the prototypical tMSP PSR J1023+0038 based on VLBA radio astrometry \citep[$1.37 \pm 0.04 $ kpc; ][]{Deller2012}, IGR J18245-2452 from its position within the globular cluster M28 \citep[5.5 kpc; ][(2010 edition)]{Harris1996}, the tMSP candidate 3FGL J1544.6–1125 based on optical modeling of the companion \citep[$3.8 \pm 0.7$ kpc; ][]{J1544_orbital}, and the tMSP candidate 3FGL J0427.9–6704 based on its Gaia DR3 parallax \citep[$2.9^{+0.7}_{-0.5}$ kpc; ][]{Lindgren2021}. In the following distance estimates based on tMSP X-ray luminosity we exclude 3FGL J0427.9-6704 due to its flare-dominated X-ray behavior, and we consider the three Galactic field sources for comparisons to other properties.

Like many tMSPs, J0640A does not have a high-significance Gaia parallax measurement. The mean X-ray luminosity of tMSPs in their disk-state is of order $L_X \sim 10^{33}$ erg s$^{-1}$. If we consider the range of $L_X = (2-8) \times 10^{33}$ erg s$^{-1}$ observed in tMSPs and candidates for the mean 1--10 keV X-ray luminosity of J0640A, then a distance range of 6.6--13.3 kpc is possible for this source. Similarly, if we consider the $L_{\gamma} = 6-30 \times 10^{33}$ erg s$^{-1}$ range of $\gamma$-ray luminosities seen for PSR J1023+0038 and tMSP candidate 3FGL J1544.6–1125, then J0640A must lie at a distance of 4.9--11.1 kpc. \cite{J0407_discovery} also pointed out that the absolute Gaia $BP$ magnitude of observed tMSPs has a narrow range, between 5.7--6.3 mag for the three Galactic field tMSPs described above. The corresponding distance range for J0640A would be 7.6--10.0 kpc, overlapping with the distance ranges estimated from the X-ray and $\gamma$-ray luminosities.

With a likely distance $>4.9$ kpc, the high proper motion of J0640A implies a transverse velocity $>313$ km s$^{-1}$, placing this system at the high end of the observed MSP 2D space velocity distribution but lower than the transverse velocity of candidate tMSP 3FGL J1544.6–1125 \citep[$v_t = 424 \pm 78$ km s$^{-1}$; ][]{Strader2019_redback_demographics}. The highest transverse velocity measured for a MSP is $\sim300$ km s$^{-1}$ for the wide-binary PSR J1024--0719 \citep{J1024:Bassa:2016, msp_astrometry:Ding:2023}. The velocity distribution of MSPs both isolated and binary is expected to be primarily determined by their evolution in the Galactic potential rather than by their natal kicks \citep{kicks:ODoherty:2023, velocity_evolution:Disberg:2024}. Velocities $\gg300$ km s$^{-1}$ are unlikely, so if J0640A is a tMSP, then this system is probably located at the low end of the allowed distance range to be consistent with the luminosity of tMSPs.

With a $2.4\sigma$ Gaia DR3 parallax measurement, the distance to J1824A is somewhat more constrained. The median distance is 1.5 kpc, with a large $2\sigma$ range of 0.8--6.4 kpc. Based on the Swift flux and Gaia distance range, its mean 1--10 keV X-ray luminosity is within the range $L_X \sim 3.9 \times 10^{31}$ to $2.5\times10^{33}$ erg s$^{-1}$. Similarly, if the Fermi source is indeed associated with this binary then the 0.1--100.0 GeV $\gamma$-ray luminosity is in the range $L_{\gamma} = 0.2-13.2 \times 10^{33}$ erg s$^{-1}$. If J1824A is a tMSP, then the minimum distances to be as luminous as the faintest observed tMSP mean X-ray and $\gamma$-ray luminosities are 4.0 and 4.3 kpc respectively. To be consistent with the observed range of tMSP absolute Gaia $BP$ magnitudes, J1824A would have to lie between 5.4--7.1 kpc. A system at 4.0 kpc with the high proper motion of J1824A would have a transverse space velocity of 351 km s$^{-1}$, again high for the observed MSP space velocity distribution but not the highest observed for a tMSP candidate. Velocities in strong excess of this value are again unlikely for a MSP, so we expect J1824A to lie at the low end of the allowed distance range.

\subsection{Radio Non-detections}

At long wavelengths, most disk-state tMSPs are faint flat-spectrum radio continuum sources. Observations of the prototypical tMSP PSR J1023+0038 with the Very Large Array showed evidence for a compact synchrotron jet \citep{J1023_Deller2015}. Simultaneous radio and X-ray observations of PSR J1023+0038 and candidate tMSP 3FGL J1544.6--1125 have demonstrated radio emission that is anti-correlated with the high and low X-ray modes \citep{J1023_VLA_Chandra, J1023_Baglio2023, Gusinskaia2024}. 

PSR J1023+0038 was observed by the VLA in 2013--2014 after its most recent transition to the disk state to display a mean radio luminosity of $L_{5 GHz} = 9.7 \times 10^{26}$ erg s$^{-1}$, and reached about $L_{5 GHz} = 1.9 \times 10^{27}$ erg s$^{-1}$ during the anti-correlated low X-ray mode \citep{J1023_Deller2015}. At a relatively close distance of 1.37 kpc, the median observed flux density of the prototypical tMSP was 87 $\mu$Jy. At its brightest during a radio flare, it reached $F_{10 GHz} =$ 533 $\mu$Jy. Radio observations of candidate tMSPs find similar faint radio continuum sources: the candidate tMSP 3FGL J1544.6--1125 was observed by \cite{Gusinskaia2024} at a radio flux density of 56.6 $\mu$Jy during the X-ray high mode in one epoch of VLA observations, corresponding to $L_{5 GHz} = 4.9 \times 10^{27}$ erg s$^{-1}$ at a distance of 3.8 kpc. The candidate 3FGL J0427.9-6704 is an exception to the radio-faintness seen in other tMSPs, with a mean out-of-eclipse flux density of $F_{5.5 GHz} = 290$ $\mu$Jy while still showing a flat spectral index $\alpha = 0.07 \pm 0.07$ like other tMSPs \citep{J0427_Li2020}. With the updated distance of 2.9 kpc, this corresponds to $L_{5 GHz} = 1.5 \times 10^{28}$ erg s$^{-1}$, an order of magnitude brighter than the mean radio luminosity of PSR J1023+0038.

For a flat-spectrum radio source as luminous as PSR J1023+0038 located at 4.9 or 4.0 kpc, the lower limits of the distance ranges we suggest for J0640A and J1824A respectively, the predicted flux densities in the RACS bands with the deepest upper limits would be $F_{1.7 GHz} = 6.8$ $\mu$Jy and $F_{0.9 GHz} =  10.2$ $\mu$Jy. These are well below our measured upper limits for each source in the shallow RACS and VLA survey imaging that is available. Sources at 4.9 and 4.0 kpc with the higher luminosity of 3FGL J0427.9-6704 would be observed at $F_{1.7 GHz} = 104.4$ $\mu$Jy and $F_{0.9 GHz} =  156.7$ $\mu$Jy respectively. These are closer to the measured $3\sigma$ upper limits, but still below the sensitivity of the available data for J0640A. At $2.3\sigma$ confidence, the upper limit for J1824A is not consistent with the bright radio emission case. Targeted radio continuum observations are necessary to further characterize radio emission from these sources. 

\subsection{The Space Density of tMSPs in the Galactic Field}

Since tMSPs are binary millisecond pulsars, they are likely to have similar kinematic and spatial distributions to other populations of millisecond pulsars. Even if the progenitor binaries of these systems were mostly born in the disk, simulations indicate that natal kicks and subsequent deceleration in the Galactic potential produce an extended spatial distribution (e.g., \citealt{space_distributions:Sartore:2010, velocity_evolution:Disberg:2024}), consistent with the observation that many millisecond pulsars with confident astrometry have been observed outside of the disk \citep[e.g.,][]{meerkat_astrometry:Shamohammadi:2024}. Therefore, as a rough initial estimate, we can take the local space density of tMSPs as constant.

PSR J1023+0038 is the closest and most well-studied tMSP. In its current (and former) subluminous disk state, its X-ray emission is well above the survey depth of the all-sky 4--12 keV SRG/ART-XC survey \citep{SRG/ART-XC}, in which it is well-detected \citep{Sazonov2024}. No new tMSPs have been identified among X-ray sources of similar flux, so this prototypical tMSP is likely to be the nearest disk-state system.

Assuming a constant space density of tMSPs, modeling their spatial distribution as a Poisson point process, and taking PSR J1023+0038 as the nearest tMSP, its distance of $1.37\pm0.04$ kpc \citep{Deller2012} implies a local space density of $6.4^{+10.7}_{-4.8} \times 10^{-2}$ kpc$^{-3}$, corresponding to $137^{+230}_{-103}$ tMSPs within 8 kpc. If we make the more uncertain but still reasonable assumption that the other confirmed field tMSP XSS J1227--4859 is the second-nearest at about 2.0 kpc, as suggested by both its Gaia DR3 parallax \citep{GaiaCollaboration2023,Lindgren2021} and by light curve modeling \citep{J1227:deMartino:2014}, then the space density inferred is $5.0^{+4.9}_{-2.9} \times 10^{-2}$ kpc$^{-3}$, consistent with but somewhat better constrained than that inferred from PSR J1023+0038. The tMSP population within 8 kpc implied by this density is $107^{+105}_{-62}$. As the density of tMSPs is likely to increase at least somewhat toward the center of the Galaxy, this estimate is probably an undercount of the total population; if PSR J1023+0038 and XSS J1227--4859 are more luminous in the disk state than typical tMSPs, this would also lead to an undercount. Although the distances for many Galactic field candidate tMSPs are not well constrained, they are all likely to be $\lesssim 8$--10 kpc, suggesting that the 9 known candidates represent $\lesssim 10\%$ of the population, and it is reasonable to think there are yet-undiscovered tMSPs at distances of only 3--4 kpc.

\subsection{Future Work}

While tMSPs were originally defined by direct evidence for transition between rotation-powered and accretion-powered states, the long durations without recent transitions in the confirmed systems ($\sim12$ yr for PSR J1023+0038; $\sim$13 yr for XSS J12270-4859) and abundance of candidate systems without observed transitions indicate that the distribution of transition times may have more weight at decades or longer timescales than was perhaps thought earlier in their discovery. Expanding the sample of tMSPs remains a priority to characterize the transition timescales and distinguish the properties that are features of the class rather than features of individual systems. Within the population of now nine likely Galactic field tMSPs, distances and orbital periods are not well constrained for six, and several are too faint for current instruments to characterize the presence of moding. Future X-ray surveys will expand the sample to include more systems where characterizing these parameters is possible and increase the likelihood of observing a future state transition. 

The two candidate tMSPs presented here can be better understood with upcoming data releases and observations. The future release of Gaia DR4 will likely improve the parallax measurement of J1824A and possibly suggest a distance range for J0640A, allowing improved constraints on the luminosity of each source. Improved distance measurements for all of the candidate tMSPs can allow more sophisticated modeling of the spatial distribution and density of the population. While in the disk state, a targeted X-ray observation of J1824A with XMM is necessary to fully characterize the source, and may reveal the presence of flaring or bimodal state-changing behavior observed in other tMSPs. To characterize the binary orbital properties, a longer baseline of high-cadence photometry would be useful for both sources, although the optical signatures of orbital motion may be obscured until a transition to the disk-free state as was the case for XSS J12270--4859 \citep{Bassa2014}. Deep radio observations are necessary to determine if jets and material ejections similar to those of PSR J1023+0038 are present, and proposed surveys with $\mu$Jy sensitivity like the Deep Synoptic Array 2000 \citep{DSA-2000} would allow radio emission to be used alongside X-ray and optical surveys to identify new disk-state tMSPs. Simultaneous multiwavelength observing campaigns of tMSPs and strong candidates have proven necessary to study the full picture of their emission and variability, particularly when X-ray moding is detected. J0640A and J1824A should be targeted by similar joint observing campaigns to fully utilize these sources to further our understanding of tMSPs. 

Constraining the presence of pulsations at the neutron star rotation period in the X-ray, UV and optical bands is of particular interest. Combined with evidence for outflows from PSR J1023+0038 \citep{j1023:baglio:2019}, millisecond pulsations detected in these bands during the X-ray flaring and high mode at the relatively low X-ray luminosity of the tMSP disk state test the limits of accretion onto neutron stars at low accretion rates \citep{j1023:xray_pulsation:archibald2015, j1023:optical_pulsation:zampieri2019,j1023:uv_pulsation:jaodand2021}. The available data for J0640A and J1824A are not sufficient to constrain X-ray pulsations in these systems. Next-generation X-ray facilities like the Advanced X-ray Imaging Satellite \citep{AXIS:co-snr} are needed to constrain the presence of X-ray pulsations at the neutron star spin period in these faint systems as well as in the low mode of PSR J1023+0038, characterize moding in faint systems as even more are discovered, and expand the search for these rare systems to dense regions. 

\vspace{5mm}
We thank the anonymous referee for useful comments that improved the clarity of the paper. We thank the Neil Gehrels Swift Observatory Time Allocation Committee for the ToO observation of J1824A. R.K. thanks Huei Sears for detailed discussions to improve visualizations in the paper. We acknowledge support from NSF grant AST-2205550 and NASA grants 80NSSC23K1350 and 80NSSC22K1583.

Based on observations obtained at the Southern Astrophysical Research (SOAR) telescope, which is a joint project of the Ministério da Ciência, Tecnologia e Inovações (MCTI/LNA) do Brasil, the US National Science Foundation’s NOIRLab, the University of North Carolina at Chapel Hill (UNC), and Michigan State University (MSU). This work has made use of data from the European Space Agency (ESA) mission {\it Gaia} (\url{https://www.cosmos.esa.int/gaia}), processed by the {\it Gaia} Data Processing and Analysis Consortium (DPAC, \url{https://www.cosmos.esa.int/web/gaia/dpac/consortium}). Funding for the DPAC has been provided by national institutions, in particular the institutions participating in the {\it Gaia} Multilateral Agreement. 

This research has made use of data obtained from the 4XMM XMM-Newton Serendipitous Source Catalog and the 4XMM XMM-Newton Serendipitous Stacked Source Catalog compiled by the 10 institutes of the XMM-Newton Survey Science Centre selected by ESA. This work made use of data supplied by the UK Swift Science Data Centre at the University of Leicester. This research has made use of data obtained from the Chandra Data Archive provided by the Chandra X-ray Center (CXC), contained in \dataset[DOI: 10.25574/cdc.353]{https://doi.org/10.25574/cdc.353}. This publication makes use of data products from the Wide-field Infrared Survey Explorer, which is a joint project of the University of California, Los Angeles, and the Jet Propulsion Laboratory/California Institute of Technology, funded by the National Aeronautics and Space Administration. 

This work is based on data from eROSITA, the soft X-ray instrument aboard SRG, a joint Russian-German science mission supported by the Russian Space Agency (Roskosmos), in the interests of the Russian Academy of Sciences represented by its Space Research Institute (IKI), and the Deutsches Zentrum für Luft- und Raumfahrt (DLR). The SRG spacecraft was built by Lavochkin Association (NPOL) and its subcontractors, and is operated by NPOL with support from the Max Planck Institute for Extraterrestrial Physics (MPE). The development and construction of the eROSITA X-ray instrument was led by MPE, with contributions from the Dr. Karl Remeis Observatory Bamberg \& ECAP (FAU Erlangen-Nuernberg), the University of Hamburg Observatory, the Leibniz Institute for Astrophysics Potsdam (AIP), and the Institute for Astronomy and Astrophysics of the University of Tübingen, with the support of DLR and the Max Planck Society. The Argelander Institute for Astronomy of the University of Bonn and the Ludwig Maximilians Universität Munich also participated in the science preparation for eROSITA.

The national facility capability for SkyMapper has been funded through ARC LIEF grant LE130100104 from the Australian Research Council, awarded to the University of Sydney, the Australian National University, Swinburne University of Technology, the University of Queensland, the University of Western Australia, the University of Melbourne, Curtin University of Technology, Monash University and the Australian Astronomical Observatory. SkyMapper is owned and operated by The Australian National University's Research School of Astronomy and Astrophysics. The survey data were processed and provided by the SkyMapper Team at ANU. The SkyMapper node of the All-Sky Virtual Observatory (ASVO) is hosted at the National Computational Infrastructure (NCI). Development and support of the SkyMapper node of the ASVO has been funded in part by Astronomy Australia Limited (AAL) and the Australian Government through the Commonwealth's Education Investment Fund (EIF) and National Collaborative Research Infrastructure Strategy (NCRIS), particularly the National eResearch Collaboration Tools and Resources (NeCTAR) and the Australian National Data Service Projects (ANDS).

Based on observations obtained with the Samuel Oschin Telescope 48-inch and the 60-inch Telescope at the Palomar Observatory as part of the Zwicky Transient Facility project. ZTF is supported by the National Science Foundation under Grants No. AST-1440341 and AST-2034437 and a collaboration including current partners Caltech, IPAC, the Oskar Klein Center at Stockholm University, the University of Maryland, University of California, Berkeley, the University of Wisconsin at Milwaukee, University of Warwick, Ruhr University, Cornell University, Northwestern University and Drexel University. Operations are conducted by COO, IPAC, and UW.

The Pan-STARRS1 Surveys (PS1) and the PS1 public science archive have been made possible through contributions by the Institute for Astronomy, the University of Hawaii, the Pan-STARRS Project Office, the Max-Planck Society and its participating institutes, the Max Planck Institute for Astronomy, Heidelberg and the Max Planck Institute for Extraterrestrial Physics, Garching, The Johns Hopkins University, Durham University, the University of Edinburgh, the Queen's University Belfast, the Harvard-Smithsonian Center for Astrophysics, the Las Cumbres Observatory Global Telescope Network Incorporated, the National Central University of Taiwan, the Space Telescope Science Institute, the National Aeronautics and Space Administration under Grant No. NNX08AR22G issued through the Planetary Science Division of the NASA Science Mission Directorate, the National Science Foundation Grant No. AST-1238877, the University of Maryland, Eotvos Lorand University (ELTE), the Los Alamos National Laboratory, and the Gordon and Betty Moore Foundation.
\vspace{5mm}
\facilities{SOAR, Fermi-LAT, Swift (XRT), XMM-Newton, Chandra, eROSITA, Gaia, ZTF, Pan-STARRS}
\software{Astropy \citep{astropy:2013, astropy:2018, astropy:2022}, specutils \citep{specutils}, IRAF \citep{IRAF}, CARTA \citep{CARTA}}
\bibliography{bib}
\bibliographystyle{aasjournal}
\end{document}